%% file: main.tex
\documentclass[11pt,a4paper]{article}

\input{Packages_Macros}

\title{Probing the Cosmological Principle\\ with weak lensing shear}

\author[a]{James Adam,}
\author[a,b,c]{Roy Maartens,}
\author[d,a]{Julien Larena,}
\author[e,a]{\\ Chris Clarkson}

\affiliation[a]{Department of Physics \& Astronomy, University of the Western Cape, \\
Cape Town 7535, South Africa}
\affiliation[b]{Institute of Cosmology \& Gravitation, University of Portsmouth, \\
Portsmouth PO1 3FX, United Kingdom}
\affiliation[c]{National Institute for Theoretical \& Computational Sciences (NITheCS), \\
Cape Town 7535, South Africa}
\affiliation[d]{Laboratoire Univers et Particules de Montpellier, CNRS \& Université de Montpellier, \\
Parvis Alexander Grothendieck, Montpellier, France 34090}
\affiliation[e]{Department of Physics \& Astronomy, Queen Mary University of London,\\
London E1 4NS, United Kingdom}

\emailAdd{james.g.adam1997@gmail.com} 
\date{\today}

\abstract{The Cosmological Principle is a cornerstone of the standard model of cosmology and shapes how we view the Universe and our place within it. It is imperative, then, to devise multiple observational tests which can identify and quantify possible violations of this foundational principle. One possible method of probing large-scale anisotropies involves the use of weak gravitational lensing. We revisit this approach in order to analyse the imprint of late-time anisotropic expansion on cosmic shear. We show that the cross-correlation of shear $E$- and $B$-modes on large scales can be used to constrain the magnitude (and possibly direction) of anisotropic expansion. We estimate the signal to noise for multipoles $10\lesssim \ell\lesssim 100$ that is achievable by a Euclid-like survey. Our findings suggest that such a survey could detect the $E$-$B$ signal for reasonable values of the late-time anisotropy parameter.}

\begin{document}

\maketitle

%%%%%%%%%%%%%%%%%%%%%%%%%%%%%%%%%%%%%%%%%%%%%%%%%%%%%%%%%%%%%%%%%%%%%%%%%%%%%%%%%%%%%%%%%%
\newpage
\section{Introduction}

A key assumption that underlies the standard model of cosmology is that, on sufficiently large scales, the universe is both homogeneous and isotropic --
{the Cosmological Principle}. Because of its numerous and far-reaching implications, this crucial assumption has deservedly been subjected to a litany of observational tests {(see e.g. \cite{Aluri:2022hzs,Euclid:2022ucc})}.

The {cosmic microwave background} (CMB) provides tight constraints of $\sim 10^{-5}$ on anisotropy around the time of recombination. The rapid decay of un-sourced anisotropic expansion  from recombination to today yields the shear constraint $\sim 10^{-10}$--$10^{-9}$ \cite{bunn_how_1996, kogut_limits_1997, martinez-gonzalez_delta_1995, maartens_anisotropy_1996, saadeh_how_2016}. Combining CMB data with measurements of galaxy baryon acoustic oscillations can improve this constraint by several orders of magnitude \cite{akarsu_testing_2023}. Additionally, a period of inflation rapidly isotropises the Universe and severely diminishes any primordial anisotropy \cite{anninos_how_1991,pitrou_predictions_2008}. These considerations, along with observations of the relative abundances of light elements \cite{rothman_effects_1982, rothman_nucleosynthesis_1984, matzner_conjecture_1986, campanelli_helium-4_2011}, appear to disfavour a period of significant anisotropic expansion during the early evolution of the Universe.

Nevertheless, these considerations do not completely rule out the possibility of a period of anisotropic expansion. In particular, {\em late-time anisotropy} driven by dark energy with intrinsic anisotropic pressure is quite compelling, since it has the potential to generate deviations from isotropy at late times when dark energy begins to dominate the Universe's energy content. Anisotropic pressure appears in models of magnetised dark energy \cite{barrow_limits_1997,sharif_dynamics_2010} as well as dark energy possessing an anisotropic equation of state \cite{koivisto_accelerating_2008, koivisto_anisotropic_2008, akarsu_lrs_2010, akarsu_bianchi_2010, akarsu_sitter_2010, kumar_bianchi_2012,appleby_probing_2013, akarsu_scalar_2020}. Moreover, the presence of effective anisotropic stresses is a fairly general feature of modified theories of gravity \cite{mukhanov_theory_1992,damour_non-linear_2002,kunz_dark_2007, saltas_anisotropic_2011}. 

Measurements of late-time anisotropy and anisotropic expansion have  primarily focused on probes of the Hubble diagram using type Ia supernovae (SN\,Ia) data \cite{kolatt_constraints_2001, jain_search_2007, cooray_measuring_2010, brunthaler_vlbi_2010, colin_probing_2011, cai_direction_2012, feindt_measuring_2013, bahr-kalus_constraints_2013, appleby_testing_2014, schucker_bianchi_2014, wang_probing_2014, appleby_probing_2015, jimenez_anisotropic_2015, javanmardi_probing_2015, ghodsi_supernovae_2017, 
andrade_isotropy_2018, andrade_model-independent_2018, soltis_percent-level_2019, colin_evidence_2019, salehi_are_2020, mohayaee_supernovae_2021, cowell_potential_2023, rahman_new_2022,Kalbouneh:2022tfw, Maartens:2023tib,Kalbouneh:2024szq,Kalbouneh:2024yjj}. {Tests of isotropy have also been performed to check whether the rest-frame of matter coincides with that of the CMB, as required by the Cosmological Principle \cite{1984MNRAS.206..377E}. There are many claims of an inconsistency in radio continuum and quasar samples (see e.g. \cite{Aluri:2022hzs}), but  a recent analysis of the eBOSS surveys finds consistency \cite{daSilveiraFerreira:2024ddn}.}

A lesser-known probe of late-time anisotropy is weak gravitational lensing \cite{amendola_measuring_2008}. The presence of $B$-modes {on large scales} in the weak-lensing signal is indicative of a violation of isotropy \cite{pitrou_weak_2013}. This contrasts starkly with standard weak-lensing wisdom which, being built on perturbed Friedmann\hyp{}Lema\^{i}tre\hyp{}Robertson\hyp{}Walker (FLRW) spacetimes, predicts that $B$-mode shear should not be generated by large-scale structure at significant levels \cite{schneider_weak_2006, bartelmann_weak_2016, kitching_limits_2017}. On the contrary, detection of these odd-parity shear patterns at significant levels is usually seen as a sign of non-linearities \cite{bernardeau_full-sky_2010,cooray_second_2002, hilbert_ray-tracing_2009, schneider_b-modes_2002}, or even systematic effects that need to be removed \cite{crittenden_discriminating_2002,bartelmann_weak_2016,troxel_intrinsic_2014,chisari_intrinsic_2015,codis_intrinsic_2015}. {On large scales where late-time anisotropy is probed, the non-linear effects will not be a problem, but systematics can arise on very large scales.} 
Nevertheless, once all systematic sources of error and spurious signals are eliminated, the detection of large-scale $B$-modes at a particular magnitude would allow us to place bounds on large-scale anisotropic expansion in the late Universe. 

The fundamental object in gravitational lensing is the {Jacobi matrix} $\D$ which maps from observed angular size $\bs{\theta}_O$ to physical separation at source $\bs{\xi}_S$
\begin{equation}
    \bs{\xi}_S = \D \bs{\theta}_O.
\end{equation}
The Jacobi matrix is usually decomposed into convergence $\kappa$, rotation $\psi$, and shear $\bs{\gamma}$ as 
\begin{equation}
    \D = d_A \left[ \mqty(1-\kappa & 0 \\ 0 & 1-\kappa) + \mqty(0 & -\psi \\ \psi & 0) + \mqty(-\gamma_1 & \gamma_2 \\ \gamma_2 & \gamma_1)\right],
\end{equation}
where $d_A$ is the background FLRW angular diameter distance. One can show that the rotation angle $\psi \sim \gamma^2$ can be safely ignored in the weak-lensing regime where $\gamma \ll 1$. Being a scalar object on the celestial sphere, the convergence can be expanded in terms of spherical harmonics
\begin{equation}
    \kappa(\vb{n}) = \sum_{\ell, m} \kappa_{\ell m} Y_{\ell m}(\vb{n}),
\end{equation}
where the observation direction $\vb{n}$ specifies a point on the sphere. The shear, on the other hand, is a spin-2 tensor variable and thus needs to be expanded in terms of spin-weighted spherical harmonics
\begin{equation}
    \gamma_1(\vb{n})  \pm i\gamma_2(\vb{n}) = \sum_{\ell,m}(E_{\ell m} \pm i B_{\ell m}) Y^{\pm 2}_{\ell m}(\vb{n}).
\end{equation}
This decomposition separates the shear multipoles into even ($E$) and odd ($B$) parity contributions. Importantly, however, linear scalar perturbations of FLRW spacetimes generate only $E$-modes. Within the FLRW context, only vector, tensor, or higher-order scalar perturbations give rise to $B$-modes \cite{bernardeau_full-sky_2010, yamauchi_weak_2012, yamauchi_full-sky_2013}.

Building on their previous work, \cite{pitrou_weak_2013} developed a two-parameter perturbation scheme which handles small deviations from isotropic expansion in a systematic way \cite{pitrou_weak-lensing_2015} and used it to calculate the $B$-mode shear generated by the coupling of large-scale anisotropic expansion and scalar metric perturbations. Subsequently, these results were used to quantify the magnitude and behaviour of observables for surveys that will be carried out using the Euclid satellite and the Square Kilometre Array (SKA) radio telescope \cite{pereira_weak-lensing_2016}. However, due to the limited availability of cosmic shear data at the time, \cite{pereira_weak-lensing_2016} was only able to make a very rough estimate of $\sigma_0/\H_0\lesssim 0.4$ for the strength of late-time anisotropic expansion using CFHTLenS data \cite{kitching_3d_2014}. 
In addition,
the estimate by \cite{pereira_weak-lensing_2016} was made using the $E$-$B$ cross-correlation at an angular scale of $\ell\sim 2000$, where additional $B$-mode signals are generated by non-linear dynamics.

In this work, we make use of the two-fold perturbation framework of \cite{pitrou_weak_2013}, {giving details of a model for generating late-time anisotropy and carrying out a systematic analysis of the $E$-$E$, $E$-$B$ and $B$-$B$ signals. We confirm that the $B$-$B$ signal is below the noise for reasonable values of the anisotropy parameter. Then we consider the larger $E$-$B$ power, exploiting the tomography  of a Euclid survey \cite{blanchard_euclid_2020, deshpande_euclid_2024}. We make a careful estimate of the signal to noise, using a conservative range of multipoles that avoids the very largest scales $\ell\lesssim 10$, where difficult systematics typically arise, while staying in the large-scale regime $\ell\lesssim 100$, in order to avoid non-linear contamination.  We make use of Halofit \cite{smith_stable_2003, takahashi_revising_2012, bird_massive_2012} in order to account for (mild) non-linearities in observed signals and to determine a cut-off scale $\ell_{\text{max}}$ which avoids any possible small-scale $B$-mode effects. We construct a simple statistical estimator 
(\autoref{eqn:P_lM_Estimator}) which contains information about the magnitude and direction of anisotropic expansion. We then estimate the signal-to-noise ratios of this observable for the Euclid tomographic photometric survey, concluding that the signal should be detectable for anisotropy parameters that are consistent with current constraints.}

\subparagraph{Notation:} We work in units where $c=1$ and the metric has signature $-+++$. Spacetime indices run from 0 to 3 and are represented by lower-case Greek letters, while lower-case Latin letters $i,j,\ldots$ denote spatial indices. These indices are also used to label the ten Euclid tomographic redshift bins, but there should be no confusion, since the two types of index are never used in the same expressions simultaneously. Upper-case Latin letters  index the two angular directions associated with screen space.

%%%%%%%%%%%%%%%%%%%%%%%%%%%%%%%%%%%%%%%%%%%%%%%%%%%%%%%%%%%%%%%%%%%%%%%%%%%%%%%%%%%%%%%%%%

\section{Modelling anisotropic expansion}

The standard $\Lambda$CDM model is built upon the framework of a perturbed FLRW universe. Since it is constructed to obey the Cosmological Principle, the general FLRW metric (i.e. possibly spatially curved) possesses spatial slices that are both isotropic and homogeneous. If one relaxes the constraint of isotropy then one arrives at the so-called Bianchi models of spacetime. The simplest of these models -- Bianchi-I -- describes a universe which expands along three orthogonal spatial directions.

For the purposes of clarity and illustration, we restrict ourselves to an axisymmetric Bianchi-I model, for which two of the expansion rates are equal. Nevertheless, most of the equations we present do not rely upon this assumption.

    \subsection{Metric and shear}  
A Bianchi-I universe is described by a spatially-Euclidean, homogeneous, and anisotropic metric. This simple anisotropic geometry enjoys three orthogonal spatial Killing vectors ${\partial}_i$. In co-ordinates in which the spatial axes are aligned with these directions of symmetry, the general form of the line element is given by
\begin{equation}\label{eqn:Scale_Fact_First}
	\dd s^2 = -\dd t^2+a^2(t)\gamma_{ij}(t)\dd x^i\dd x^j = a^2(\eta)\big[-\dd \eta^2+\gamma_{ij}(\eta)\dd x^i\dd x^j\big],
\end{equation}
where cosmic time $t$ and conformal time $\eta$ are related in the usual manner: $a(\eta)\dd \eta=\dd t$. The spatial metric $\bs{\gamma}$ is defined as (no sum over $i$)
\begin{equation}\label{eqn:beta_i_first}
	\gamma_{ij}(t) =\exp[2\beta_i(t)]\delta_{ij}, \quad \gamma^{ij}(t) =\exp[-2\beta_i(t)]\delta^{ij}.
\end{equation}
The three $\beta_i$ functions are subject to the constraint $\sum_{i}\beta_i(t)=0$. Clearly, if $\beta_1(t)=\beta_2(t)=0$ (up to a constant) we recover the FLRW spacetime. If the $\beta_i$ are not equal to one-another, however, we will have anisotropic expansion (or contraction). In analogy to the conformal Hubble parameter $\H \equiv a'/a$, which quantifies isotropic expansion, the (conformal) {geometric shear} $\bs{{\sigma}}$ is a measure of the rate of anisotropic expansion. This quantity is defined as
\begin{equation}
	{\sigma}_{ij}\equiv \frac{1}{2}{\gamma}_{ij}'={\beta_i}'\gamma_{ij},
\end{equation}
where the prime represents a derivative with respect to conformal time $\eta$. It is clear that $\bs{{\sigma}}$ is both symmetric and traceless (i.e. $\gamma^{ij}{\sigma}_{ij}=\sum_{i}{\beta}_i'=0$). Furthermore, we define the (squared) amplitude of the shear to be
\begin{equation}
    \sigma^2\equiv\sigma_{ij}\sigma^{ij}=\sum_{i}\beta_i'^2.
\end{equation}

If we impose axisymmetry and choose to align the axis of symmetry with the $z$-direction, the expansion coefficients and components of the shear simplify to 
\begin{equation}
    \beta_1=\beta_2=-\frac{1}{2}\beta_3, \quad  \sigma_1 = \sigma_2 = -\frac{1}{2}\sigma_3,
\end{equation}
where $\Big(\sigma\indices{^i_j}\Big) = \text{diag}(\sigma_1,\sigma_2,\sigma_3)$ and $\sigma_i=\beta_i'$.

        \subsection{Equations of motion}
The momentum constraint of the Einstein equations precludes the possibility of a net momentum density/energy flux in Bianchi-I spacetimes. Thus, the most general energy-momentum tensor that is consistent with the restrictions of a Bianchi-I geometry is 
\begin{equation}\label{eqn:Gen_EM}
	T^{\mu\nu} = (\rho+P)u^{\mu}u^{\nu}+Pg^{\mu\nu}+\Pi^{\mu\nu}.
\end{equation}
The energy density $\rho$, isotropic pressure $P$, and fluid velocity $\bs{u}$ are analogous to their perfect-fluid counterparts. One can easily confirm that $\bs{u}=a^{-1}\partial_{\eta}$ is a geodesic in Bianchi-I and hence that we can work in a frame in which the fluid velocity $u^{\mu}=a^{-1}\delta\indices{^\mu_0}$. The anisotropic stress tensor is symmetric ($\Pi_{\mu\nu}=\Pi_{\nu\mu})$, traceless ($\Pi\indices{^\mu_\mu}=0$), and transverse ($u_{\mu}\Pi^{\mu\nu}=0$).

The Einstein field equations for a Bianchi-I metric take the form \cite{pitrou_theory_2007}
\begin{subequations}\label{eqn:BI_EoM}
\begin{align}
		\H^2 &= \frac{8\pi G}{3} a^2\rho+\frac{1}{6} \sigma^2\label{eqn:Friedmann_Bianchi_I} \\
		\H'  &= -\frac{4\pi G}{3} a^2(\rho+3P)-\frac{1}{3}\sigma^2 \\
		(\sigma\indices{^i_j})' &= -2\H\sigma\indices{^i_j}+ 8\pi G a^2\tilde{\pi}\indices{^i_j} \quad  
{\mbox{where}\quad a^2\tilde{\pi}_{ij}\equiv \Pi_{ij}} ,
  \label{eqn:Shear_EoM}	
\end{align}
\end{subequations}
while the general conservation equation reads
\begin{equation}\label{eqn:General_Matter_Conservation}
		\rho' = -3\H(\rho+P)-\sigma^{ij}\tilde{\pi}_{ij}  , 
\end{equation}
for each species of fluid. Note that we have defined $\tilde{\pi}_{ij}$ so that its indices can be raised (lowered) by $\gamma^{ij}$ ($\gamma_{ij}$). In the above equations, the total energy density $\rho = \rho_m+\rho_{\gamma}+\rho_{de}$ is the sum of the non-relativistic matter (cold dark matter and baryons), radiation, and dark energy densities, respectively. The non-relativistic matter and radiation possess barotropic equation of state parameters $w_m=0$ and $w_\gamma=1/3$, and vanishing anisotropic stress $\tilde{\pi}^{m}_{ij}=\tilde{\pi}^{\gamma}_{ij}=0$. Furthermore, in order to close this set of equations one needs to specify an equation of state for the pressure $P_{de}$ as well as a model for the anisotropic stress term $\tilde{\pi}_{ij}$.

Defining the density parameters
\begin{equation}
	\Omega_\rho \equiv \frac{8\pi G a^2 \rho}{3\H^2}, \qquad \Omega_\sigma \equiv \frac{\sigma^2}{6\H^2},
\end{equation}
allows us to write the Friedmann constraint \autoref{eqn:Friedmann_Bianchi_I} as
\begin{equation}
 	\Omega_m+\Omega_{de}+\Omega_\gamma+\Omega_\sigma = 1.
 \end{equation}
The quantities $\Omega_m$, $\Omega_{de}$ and $\Omega_\gamma$ have been measured to sub-percent precision by the Planck survey \cite{aghanim_planck_2020}. This allows us to place bounds on the value of $\sigma/\H$ and hence quantify the possible amount of anisotropic expansion. Now, we know that on large scales the universe is well-described by the isotropic $\Lambda$CDM model (i.e. cold or non-relativistic dark matter with a cosmological constant $\Lambda$). For an anisotropic model to be consistent with observations, we must therefore demand that the ratio $\sigma/\H$ be very small over the history of the universe. In other words, our consistency condition is
\begin{equation}\label{eqn:Small_Shear_Consistency}
	\frac{\sigma}{\H} \ll 1.
\end{equation}

The current values of the matter (baryonic and cold dark) and dark energy density parameters obtained by the Planck mission in a six-parameter $\Lambda$CDM model fit are $\Omega_{m0} = 0.315(7) $ and $\Omega_{de0} = \Omega_{\Lambda 0} = 0.685(7)$, respectively \cite{workman_review_2022,aghanim_planck_2020}. If we assume that the shear density parameter $\Omega_{\sigma}$ is smaller than the uncertainty brackets surrounding $\Omega_{m0}$ and $\Omega_{de0}$, we can obtain a crude estimate for a consistent value of $\Omega_{\sigma 0}$. The shear density parameter can be thought of as `hiding' in the uncertainties of the other density parameters. In other words, we set the constraint $\Omega_{\sigma 0}\lesssim 10^{-2}$. In the case of  axisymmetry, this translates to $|\flatfrac{(\sigma_1)_0}{\H_0}| \lesssim 10^{-1}$.

    \subsection{Equation of state and anisotropic stress model}

%[{\bf RM}: Rather make this a new section 4? Same for sec. 3.3 and 3.4?]
Without a mechanism to drive it, the shear decays as $\sigma \propto a^{-2}$ and any effects of anisotropic expansion quickly become negligible. One possible source of late-time anisotropic expansion is a type of anisotropic dark energy. Following \cite{pitrou_weak-lensing_2015}, we assume that the matter in the universe is described by a pressureless fluid (perfect fluid with $w_m=0$) and a dark energy component with anisotropic pressure.

In order to close the Einstein \autoref{eqn:BI_EoM} and conservation \autoref{eqn:General_Matter_Conservation} system, both the pressure and anisotropic stress of the dark fluid need to be modelled or given an equation of state. The specific form of this model can, in principle, mimic the expected anisotropic stresses generated by some physical process. For our purposes, we consider a toy model inspired by \cite{pereira_weak-lensing_2016}
\begin{equation}
 	\Pi\indices{^i_j}(a) = f(a)W\indices{^i_j},
\end{equation}
where $W\indices{^i_j}$ is a constant matrix and $f$ is a time-dependent function which has the dimensions of energy density. For models of this type, the formal solution to \autoref{eqn:Shear_EoM} is 
\begin{equation}
    \sigma\indices{^i_j}(a) = \left(\frac{a}{a_I}\right)^{-2}\left[{\sigma_I}\indices{^i_j}+8\pi G a_I^2\int_{a_I}^a\frac{\dd {\bar{a}}}{a_I} \left(\frac{{\bar{a}}}{a_I}\right)^{3} \frac{f({\bar{a}})}{\H({\bar{a}})}W\indices{^i_j}\right].
    \label{eqn:Shear_Formal_Soln}
\end{equation}
where $a_I$ is the value of the scale factor at some time $t_I$, and ${\sigma_I}\indices{^i_j} = \sigma\indices{^i_j}(a_I)$. The solution in \autoref{eqn:Shear_Formal_Soln} contains a decaying mode ($\sim a^{-2}$) and a possible growing mode (the integral) which correspond, respectively, to the homogeneous and particular solution of the differential \autoref{eqn:Shear_EoM}. The growing mode is responsible for late-time anisotropic expansion while the decaying mode should become negligible during the early evolution of the Universe. If we wish to constrain the value of the spatial shear today, we therefore need to consider carefully how to select only the growing mode in a numerical implementation of \autoref{eqn:Shear_Formal_Soln} as this solution is highly sensitive to model parameters. In Appendix \ref{sec:Stress_Model_Parameters} we detail how we use this solution to select model parameters $W\indices{^i_j}$ which carefully avoid the decaying mode.

We use a particularly simple form for $f$ which encourages shear growth at late times:
\begin{equation}
    f(a) = \frac{\H_0^2}{8\pi G} \Omega_{de}(a),
\end{equation}
with the dark energy density parameter  $\Omega_{de} \equiv 8\pi G a^2\rho_{de}/\H^2$. The evolution of the dark energy density depends on the behaviour of the shear and anisotropic stress. For a constant equation of state parameter $w_{de} \equiv P_{de}/\rho_{de}$, the analytic solution is of the form
\begin{equation}
	\rho_{de}(a) = \left(\frac{a}{a_0}\right)^{-3(1+w_{de})}\left[\rho_{de0}-\int_{a_0}^{a}\frac{\dd \bar{a}}{a_0} \frac{\sigma^{ij}({\bar{a}})\tilde{\pi}_{ij}({\bar{a}})}{\H({\bar{a}})}\left(\frac{\bar{a}}{a_0}\right)^{2+3w_{de}}\right]. \label{eqn:rho_de_soln_Bianchi}
\end{equation}
A quick sanity check confirms that if $w_{de}=-1$ and the anisotropic stress is weak, then the dark energy density is approximately constant. Although we could in principle investigate other parametrisations, we fix the dark energy equation of state for the isotropic pressure $P_{de}$ to be 
\begin{equation}
    w_{de} = \frac{P_{de}}{\rho_{de}} = -1.    
\end{equation}
Importantly, we note that since we expect $|\sigma_{ij}|/\H \ll 1$ to be of the same order as $|\tilde{\pi}_{ij}|/\rho_{de}$, the product of the shear and anisotropic stress in \autoref{eqn:rho_de_soln_Bianchi} should be doubly small. Thus, in accordance with the perturbation scheme described in \S \ref{sec:Perturbation_Scheme}, we neglect this contribution and the dark energy density is approximated by a cosmological constant
\begin{equation}
    \rho_{de} \approx \rho_{\Lambda} = \frac{\Lambda}{8\pi G}.    
\end{equation}

\begin{figure}[htb]
	\centering
	\includegraphics[scale=0.6]{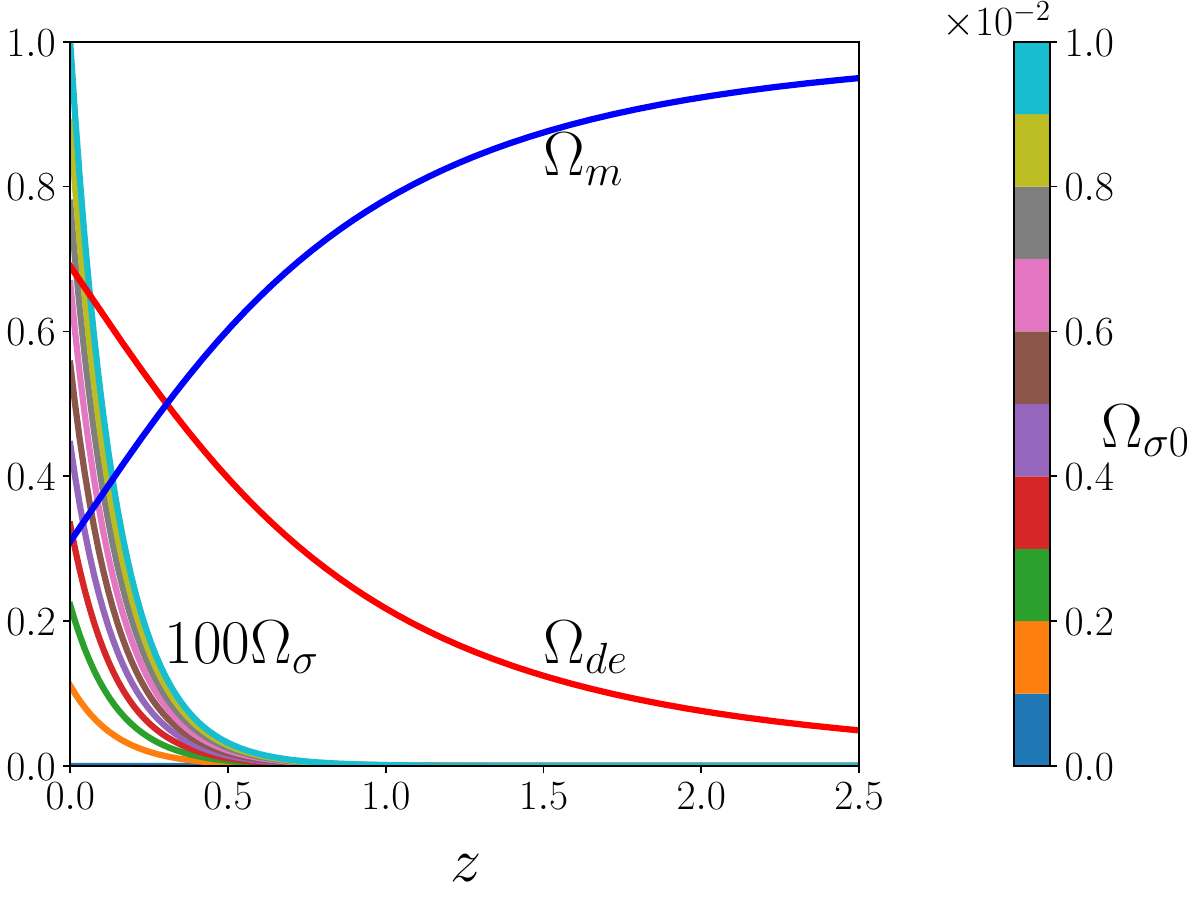 }
	\caption{Evolution of density parameters $\Omega_m$, $\Omega_{de}$, and $\Omega_{\sigma}$ (scaled by a factor of 100) up to $z=2.5$ for various values of $\Omega_{\sigma0}$.}
 \label{fig:Density_Params}
\end{figure}
The effect of varying the current shear density parameter $\Omega_{\sigma0}$ up to a maximum value of $10^{-2}$ is shown in  \autoref{fig:Density_Params}. The procedure for generating the initial conditions and model parameters necessary to compute the evolution of $\bs{\sigma}$ is outlined in Appendix \ref{sec:Stress_Model_Parameters}.

    \section{Weak shear limit and perturbation scheme}\label{sec:Perturbation_Scheme}
In principle, one could derive (as presented in \cite{pitrou_theory_2007,pitrou_predictions_2008, almeida_structure_2022}) and subsequently solve the perturbed Einstein equations for a Bianchi-I universe to predict the behaviour of universes with large-scale anisotropies. Thereafter, one could solve the Sachs equation to quantify the effect that these spacetime geometries would have on lensing observables. The effect of a background Bianchi-I universe on weak gravitational lensing has already been solved exactly in \cite{fleury_light_2015}.

It is easy to understand why the prospect of solving the perturbed Einstein equations for a Bianchi-I universe is a daunting one. Because there are fewer symmetries and more degrees of freedom at the background level, the perturbation equations are naturally more complex and do not decouple cleanly into scalar, vector, and tensor modes. Moreover, many Boltzmann codes (such as CAMB or CLASS) that solve perturbation equations as part of their repertoire are built on the backbone of FLRW cosmologies. Besides, it is well known that our Universe is described rather well by FLRW cosmologies on large scales. The natural question of treating large-scale anisotropies perturbatively has been investigated in detail by \cite{pontzen_linearization_2011} and subsequently by \cite{pitrou_bianchi_2019}. The advantage of this framework is that anisotropic cosmologies can be treated as homogeneous, deterministic (i.e. non-stochastic) perturbations atop an FLRW background and analysed accordingly. Moreover, many of the tools (both theoretical and numerical) that have been developed for traditional cosmological perturbation theory can be brought to bear. Using this idea, \cite{pitrou_weak_2013} employed a two-fold perturbation scheme in \cite{pitrou_weak-lensing_2015,pereira_weak-lensing_2016} which treats Bianchi-I degrees of freedom as a small perturbation of a spatially-flat FLRW spacetime in order to estimate $B$-mode shear generated by large-scale, late-time anisotropies.

The shear $\bs{\sigma}$ adds degrees of freedom that are not present in FLRW cosmologies. More concretely, it adds homogeneous anisotropic expansion which can lead to anisotropies on large scales. However, since the universe appears to be (mostly) well-described by an isotropic cosmology, we expect the anisotropic expansion encoded by $\bs{\sigma}$ to be much weaker than the isotropic expansion described by the Hubble rate $\H$. Thus, we work in the \emph{weak shear limit} described by \cite{pitrou_weak-lensing_2015,pereira_weak-lensing_2016}
\begin{equation}
	\left|\frac{\sigma_{ij}}{\H}\right| \ll 1.
\end{equation}
More precisely, it is assumed that $|\beta_i|\ll 1$ are small perturbations to Euclidean space and hence that the spatial metric is given by
\begin{equation}
	\gamma_{ij} \approx \delta_{ij}+2\beta_{ij},
\end{equation}
with $\beta_{ij} \equiv \text{diag}(\beta_{i})$ if our axes align with the principal directions of shear. Furthermore, we assume that
\begin{equation}
	\dv{\beta_{ij}}{\ln a} = \frac{\sigma\indices{^i_j}}{\H} 
\end{equation}
is of the same order as $\beta_{ij}$.

\begin{figure}[htb]
	\centering
	\includegraphics[scale=0.6]{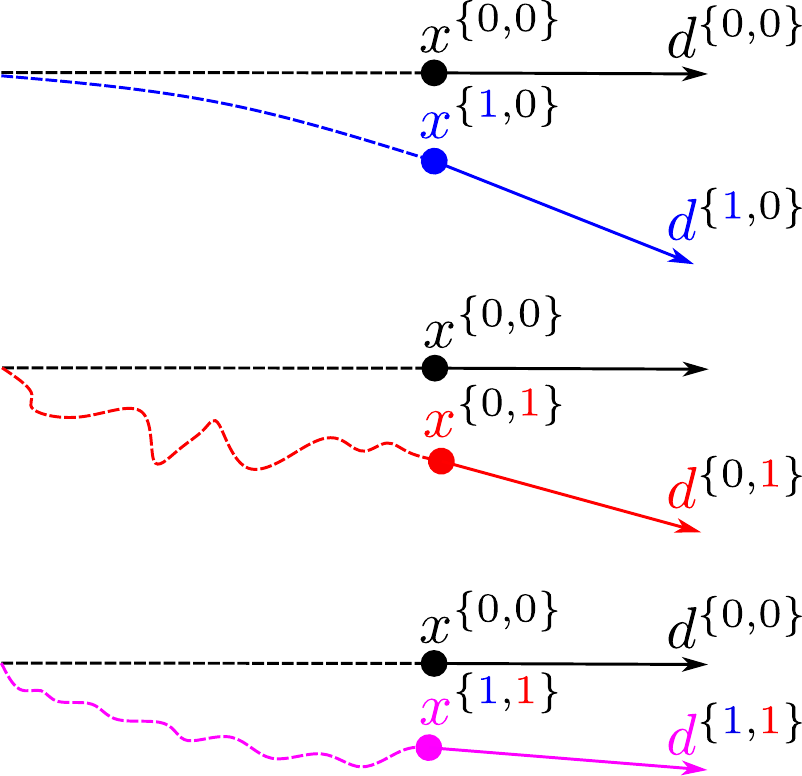}
	\caption{Diagram illustrating the perturbed path $x^{\{n,m\}}$ of a light ray and its direction vector $\bs{d}^{\{n,m\}}$ up to order ${\II}$. The perturbed quantities are colour coded, with first-order shear perturbed quantities (order ${\IO}$) being blue and first-order scalar FLRW perturbed quantities (${\OI}$) being red. The order ${\II}$ perturbed path and direction vector (which may contain products of ${\IO}$ and ${\OI}$ quantities as well as vector and tensor perturbations) are coloured magenta.}
    \label{fig:Perturbation_Scheme}
\end{figure}

This means we can treat $\bs{\sigma}$ and related quantities as small perturbations to an isotropic FLRW spacetime. These zero-mode (homogeneous) perturbations are independent of the usual scalar-vector-tensor (SVT) perturbations that are introduced upon an FLRW background when studying inhomogeneities in our universe. This suggests that we make use of a two-fold perturbation scheme which captures the order of both types of perturbations. Following  \cite{pitrou_weak-lensing_2015}, we define the perturbation order $\{n,m\}$ to correspond to a perturbed quantity of order $n$ in the shear $\bs{\sigma}$ (and its time derivatives) and order $m$ in the scalar variables (and derivatives thereof). Thus, orders $\{1,0\}$ and $\{0,1\}$ correspond to first-order perturbations in shear and scalar degrees of freedom, respectively, and quantities of the form $\left(\flatfrac{\sigma}{\H}\right)^{n}\Phi^{m}$ and $(\beta')^{n}(\partial\Psi)^{m}$ are of order $\{n,m\}$.  Vector and tensor perturbations are of at least order $\{1,1\}$ since they are dominated by scalar perturbations in FLRW cosmologies (at least in the late universe due to their rapid decay) and because they only couple to the scalar evolution equations through the shear \cite{pitrou_theory_2007,pitrou_predictions_2008}. Clearly order $\OO$ corresponds to the un-perturbed background (flat FLRW) cosmology. Written schematically, a quantity $Q$ is expanded up to order ${\II}$ as
\begin{equation}
    Q^{{\II}} = \underbrace{Q^{\OO}}_{\text{FLRW}} + \overbrace{\delta Q^{{\IO}}}^{\text{BI}} + \underbrace{\delta Q^{{\OI}}}_{\text{S}} + \overbrace{\delta Q^{{\II}}}^{\text{BI}\times\text{S} + \text{V} + \text{T}},
\end{equation}
where BI stands for Bianchi-I, S for scalar, V for vector, and T for tensor modes, respectively. For clarity, note that the object $Q^{\{n,m\}}$ contains terms of the order \emph{up to and including} $\{n,m\}$ whereas $\delta Q^{\{n,m\}}$ contains \emph{only} terms of the order $\{n,m\}$. An illustration of the two-parameter perturbation scheme applied to the path and direction of a light ray is shown in \autoref{fig:Perturbation_Scheme}. Now the problem of extracting particular perturbation orders from an expression rears its head. A general method for solving this problem is outlined in appendix \ref{sec:Arbitrary_Pert}.

The process of solving the Sachs equation in order to obtain the perturbed lensing variables up to order {\II} is long and involved. In this section, we summarise the key results of \cite{pitrou_weak_2013} and clarify the reasoning they employed to isolate the dominant contribution to the shear multipoles. More details can be found in their paper \cite{pitrou_weak-lensing_2015}. The perturbed Bianchi-I metric along with our particular gauge choice can be found in appendix \ref{sec:Perturbed_Metric}.

$E$-modes appear at all perturbation orders up to ${\II}$, while $B$-modes are only generated at order ${\II}$
\begin{subequations}
    \begin{align}
        E^{{\II}}_{\ell m} &= \delta E^{{\IO}}_{\ell m} + \delta E^{{\OI}}_{\ell m} + \delta E^{{\II}}_{\ell m}\label{eqn:E_lm_11}\\
        B^{{\II}}_{\ell m} &= \delta B^{{\II}}_{\ell m}.
    \end{align}
\end{subequations}

        \subsubsection*{Order {\IO}}
        
The order {\IO} Bianchi-I correction to the shear is given by\footnote{Note that we do not project with the source distribution $\N$ since the \IO\, correction is not stochastic and thus does not appear in the \II\, correlation functions.}
\begin{equation}
    \delta E^{{\IO}}_{\ell m}(\chi) = \sqrt{6}\left[{\mathcal{E}}_{2m}(\chi) -\frac{2}{\chi}\int_{0}^{\chi}\dd\chi_{1}\,\mathcal{E}_{2m}(\chi_1)\right]\delta_{\ell 2},
\end{equation}
where $\chi\equiv\eta_0-\eta$ is the conformal distance down the lightcone. The quadrupole coefficients $\mathcal{E}_{2m}$ are defined in terms of the homogeneous perturbation $\beta_{ij}$: 
\begin{equation}\label{eqn:Beta_2m_Multipoles}
    \mathcal{E}_{2m} = \begin{cases}
					-\sqrt{{\pi}/{5}}\,\big(\beta_{xx}+\beta_{yy}\big) & m=0\,, \\
					2 \sqrt{{\pi}/{30}}\,\big(\mp\beta_{xz}+i\beta_{yz}\big)  & m=\pm 1 \,,\\
					\sqrt{{\pi}/{30}}\,\big(\beta_{xx}-\beta_{yy} \mp 2i\beta_{xy}\big) & m=\pm 2\,. 
				\end{cases}
\end{equation}
Clearly, aligning our co-ordinate axes with the principal axes of expansion greatly simplifies the form of \autoref{eqn:Beta_2m_Multipoles}. Furthermore, imposing axisymmetry along the $z$-axis means that the $m=0$ quadrupole is the only non-zero harmonic coefficient.

        \subsubsection*{Order {\OI}}
        
At order {\OI}, the standard result expresses the $E$-mode shear in terms of the {Weyl potential} $\varphi \equiv \Phi +\Psi$
\begin{equation}\label{eqn:E_01_Multipoles}
    \delta E_{\ell m}^{{\OI}}(\chi_S) \equiv -\frac{1}{2}\bigg[ {\frac{(\ell+2)!}{(\ell-2)!}}\bigg]^{1/2}\int_{0}^{\chi_S}\dd\chi \, q(\chi,\chi_S)\varphi_{\ell m}(\chi),
\end{equation}
where the function $q$ has been defined as a weighted integral over the source distribution function $\mathcal{N}$
\begin{equation}\label{eqn:q_defn}
    q(\chi,\chi_S) = \int_{\chi}^{\chi_S}\dd\chi_1\, \frac{\chi_1-\chi}{\chi_1\chi}\mathcal{N}(\chi_1).
\end{equation}
For tomographic source distributions, we label the weight function corresponding to a tomographic bin $i$ as $q^i$, and simply replace $\N$ in \autoref{eqn:q_defn} with the source distribution $\N^i$.

The multipoles of $\varphi$ can be expressed in terms of the {transfer function} $T_\varphi$ and the {primordial curvature perturbation} $\mathcal{R}$:
\begin{equation}
    \varphi_{\ell m}(\chi) = i^{\ell}\Big({\frac{2}{\pi}}\Big)^{1/2} %\int_{\Reals[3]}
   \int \dd^{3}\vb{k}\, Y_{\ell m}^{*}(\hat{\vb{k}})j_{\ell}(k\chi) T_{\varphi}(\chi,k)\mathcal{R}(\vb{k}),
\end{equation}
where $Y_{\ell m} $ is a spherical harmonic and $j_\ell$ is a spherical Bessel function.
        \subsubsection*{Order {\II}}

The dominant contribution to the shear at order ${\II}$ is given by a post-Born correction \cite{pitrou_weak-lensing_2015}. More specifically, the Bianchi-I angular deflection $\bs{\alpha}^{{\IO}}$ couples to the scalar potential $\varphi$ to produce a source term in the Sachs equation of the form
\begin{equation}
      S^{{\II}}_{AB} =  \frac{1}{\chi} \alpha^{C{\IO}}D_C D_{\langle A}D_{B\rangle}\varphi,  
\end{equation}
which contains three covariant angular derivatives $D_A$. Each derivative generates a factor of $\mathcal{O}(\ell)$ in multipole space.
%\footnote{This is analogous to how the Fourier transform converts derivatives in real space to multiplication in Fourier space.}. 
This post-Born source term is therefore $\mathcal{O}\left(\ell^3\right)$, which should dominate other terms at sufficiently large $\ell$. With this simplification, the {\II} corrections to the $E$-modes and $B$-modes are
\begin{subequations}
    \begin{align}
        \begin{split}           
            \delta E_{\ell m}^{{\II}}(\chi_S) = \frac{(-1)^{m+1}}{4}  \largesum_{\substack{m_1 \\ \ell_2,m_2}} & \int_{0}^{\chi_S}\dd\chi\, q(\chi,\chi_S) \bigg[{\frac{(\ell_2+2)!}{(\ell_2-2)!}}\bigg]^{1/2}\,\alpha_{2 m_1}^{{\IO}}(\chi) \varphi_{\ell_2 m_2}(\chi) \\
            & \cdot \left[1 + (-1)^{\ell + \ell_2} \right]  \mqty(\ell & 2 & \ell_2\\ -m & m_1 & m_2) {^{2}F}_{\ell 2\ell_2}
        \end{split}\label{eqn:E_Mode_Multipoles}\\
        \begin{split}           
            \delta B_{\ell m}^{{\II}}(\chi_S) = \frac{(-1)^{m+1}}{4i}  \largesum_{\substack{m_1 \\ \ell_2,m_2}} & \int_{0}^{\chi_S}\dd\chi \, q(\chi,\chi_S) \bigg[{\frac{(\ell_2+2)!}{(\ell_2-2)!}}\bigg]^{1/2}\,\alpha_{2 m_1}^{{\IO}}(\chi) \varphi_{\ell_2 m_2}(\chi) \\
            & \cdot \left[1 - (-1)^{\ell + \ell_2} \right]  \mqty(\ell & 2 & \ell_2\\ -m & m_1 & m_2) {^{2}F}_{\ell 2\ell_2}.
        \end{split}\label{eqn:B_Mode_Multipoles}         
    \end{align}
\end{subequations}
The quadrupole coefficients $\alpha_{2 m}^{{\IO}}$ are defined in terms of the matrix
\begin{equation}
    \alpha_{ij}(\chi) \equiv -\frac{2}{\chi}\int_{0}^{\chi}\dd\chi_1\,\beta_{ij}(\chi_1)
\end{equation}
and follow the same pattern as \autoref{eqn:Beta_2m_Multipoles}.

We note, however, that, as with the CMB, this kind of gradient expansion is only valid for describing structures on scales larger than the typical deflection angle $\bar{\alpha}$ corresponding to a multipole $\bar{\ell}\sim \pi/\bar{\alpha}$. 

%%%%%%%%%%%%%%%%%%%%%%%%%%%%%%%%%%%%%%%%%%%%%%%%%%%%%%%%%%%%%%%%%%%%%%%%%%%%%%%%%%%%%%%%%%

\section{Angular power spectra}\label{sec:Angular_Spectra}
Under parity, the shear $E$-modes transform like multipoles of a scalar quantity, while the $B$-modes transform like those of a pseudo-scalar
\begin{equation}
    E_{\ell m} \longmapsto (-1)^\ell E_{\ell m}, \quad B_{\ell m} \longmapsto (-1)^{\ell+1} B_{\ell m}.
\end{equation}
These transformation properties have implications for the allowed angular correlation functions of shear $E$- and $B$-modes. In particular, the auto- and cross-correlation functions satisfy \cite{pitrou_weak-lensing_2015,abramo_testing_2010, samandar_cosmic_2024}
\begin{subequations}
    \begin{align}
        \expval{E_{\ell m} E_{\ell' m'}^*} &= (-1)^{\ell +\ell'} \expval{E_{\ell m} E_{\ell' m'}^*}\label{eqn:E_E_Selection_Rule} \\
        \expval{B_{\ell m} B_{\ell' m'}^*} &= (-1)^{\ell +\ell'} \expval{B_{\ell m} B_{\ell' m'}^*} \\
        \expval{E_{\ell m} B_{\ell' m'}^*} &= (-1)^{\ell +\ell'+1} \expval{E_{\ell m} B_{\ell' m'}^*}.
    \end{align}
\end{subequations}
Hence, provided that no parity-violating physics is at play, the auto-correlations vanish whenever $\ell +\ell'$ is an odd integer while the cross-correlation vanishes whenever $\ell +\ell'$ is even. These selection rules necessitate that we use statistical measures well-suited to study the off-diagonal correlators that arise frequently in studies of anisotropy.

One set of special functions that are particularly well-adapted to quantifying deviations from isotropy are {bipolar spherical harmonics} (BipoSHs). These harmonics (the properties of which are summarised in Appendix \ref{sec:BipoSH_Properties}) have been used extensively in the context of the CMB in order to investigate a wide range of possible sources of anisotropy \cite{hajian_measuring_2003, hajian_statistical_2003, hajian_cosmic_2005, ghosh_unveiling_2007, joshi_bipolar_2010, book_odd-parity_2012, aluri_novel_2015}. For two stochastic variables $X$ and $Z$, the BipoSH multipole coefficients are given by linear combinations of the raw angular correlations $\expval{X_{\ell m}Z^*_{\ell' m'}}$ \cite{hajian_cosmic_2005}
\begin{equation}\label{eqn:BipoSH_Coeff_Defn}
    ^{XZ}\!\A^{LM}_{\ell \ell'} = \sqrt{2L+1}\largesum_{m,m'}(-1)^{L+m} \mqty(\ell & \ell' & L \\ -m & m' & M) \expval{X_{\ell m}Z^*_{\ell' m'}}. 
\end{equation}
where the $2\times 3$ matrix is a Wigner 3$j$ symbol. For a statistically-isotropic Universe these coefficients reduce to the (re-scaled) diagonal power spectrum
\begin{align}
    ^{XZ}\!\A^{LM}_{\ell \ell'} = (-1)^{\ell}\sqrt{2\ell+1}\,C^{XZ}_{\ell}\delta_{\ell \ell'}\delta_{L0}\delta_{M0}.
\end{align}

Although they are a mathematically elegant measure of deviations from isotropy, it is impossible to accurately determine each BipoSH multipole individually. This is because \autoref{eqn:BipoSH_Coeff_Defn} is an invertible change of basis and we only have access to one set of multipole coefficients $X_{\ell m}$ and $Z_{\ell' m'}$ from any single all-sky map. In practice, in order to beat down cosmic variance, one has to form combinations of {\biposh} coefficients or construct an estimator based on some model of anisotropy.

Although not essential, we make use of the Limber approximation in order to compute all theoretical angular power spectra and {\biposh} coefficients. Nevertheless, this approximation should be fairly accurate for lensing observables even for $\ell \gtrsim 12$ \cite{kilbinger_precision_2017}. The non-Limber expressions for all spectra can be found in Appendix \ref{sec:Spectra_Details}.

    \subsection{Non-linear corrections}
The kernels that arise in weak lensing are broad and have a smoothing effect on the matter power spectrum. During line-of-sight projection, these kernels mix large angular modes with small spatial modes (and vice-versa). Moreover, weak lensing surveys are sensitive to redshifts $z\lesssim 3$. Consequently, non-linear features in the matter power spectrum manifest themselves at relatively large angular scales in lensing power spectra and need to be accounted for. %\note{*[Cite lensing modelling paper here]*}

$B$-modes can be generated by non-linear effects at small scales. We need to avoid scales where these additional $B$-mode mechanisms could play a significant role and overshadow the effect we are trying to study. Including some non-linear effects in our $B$-mode angular power spectra will therefore allow us to gauge the limiting scale of our analysis more effectively.

Properly accounting for non-linear features in the matter power spectrum is a complex endeavour. Nevertheless, we can make significant headway by re-scaling the scalar transfer function $T_{\varphi}$ according to \cite{challinor_weak_2006, challinor_lensed_2005}
\begin{equation}
    T_{\varphi}(z,k) \longmapsto c_{\text{NL}}(\chi,k)T_{\varphi}(z,k).
\end{equation}
The non-linear correction factor $c_{\text{NL}}$ is defined in terms of the linear and non-linear matter power spectra
\begin{equation}
    c_{\text{NL}}(z,k)^2 \equiv {\frac{P_m^{\text{NL}}(z,k)}{P_m^{\text{L}}(z,k)}}\,,
\end{equation}
and allows for a smooth interpolation between the two regimes. In this work we make use of HaloFit (through its implementation in CLASS \cite{lesgourgues_cosmic_2011-3}) in order to model the non-linear matter power spectrum $P_m^{\text{NL}}$.

\begin{figure}[htb!]
	\centering
	\includegraphics[scale=0.6]{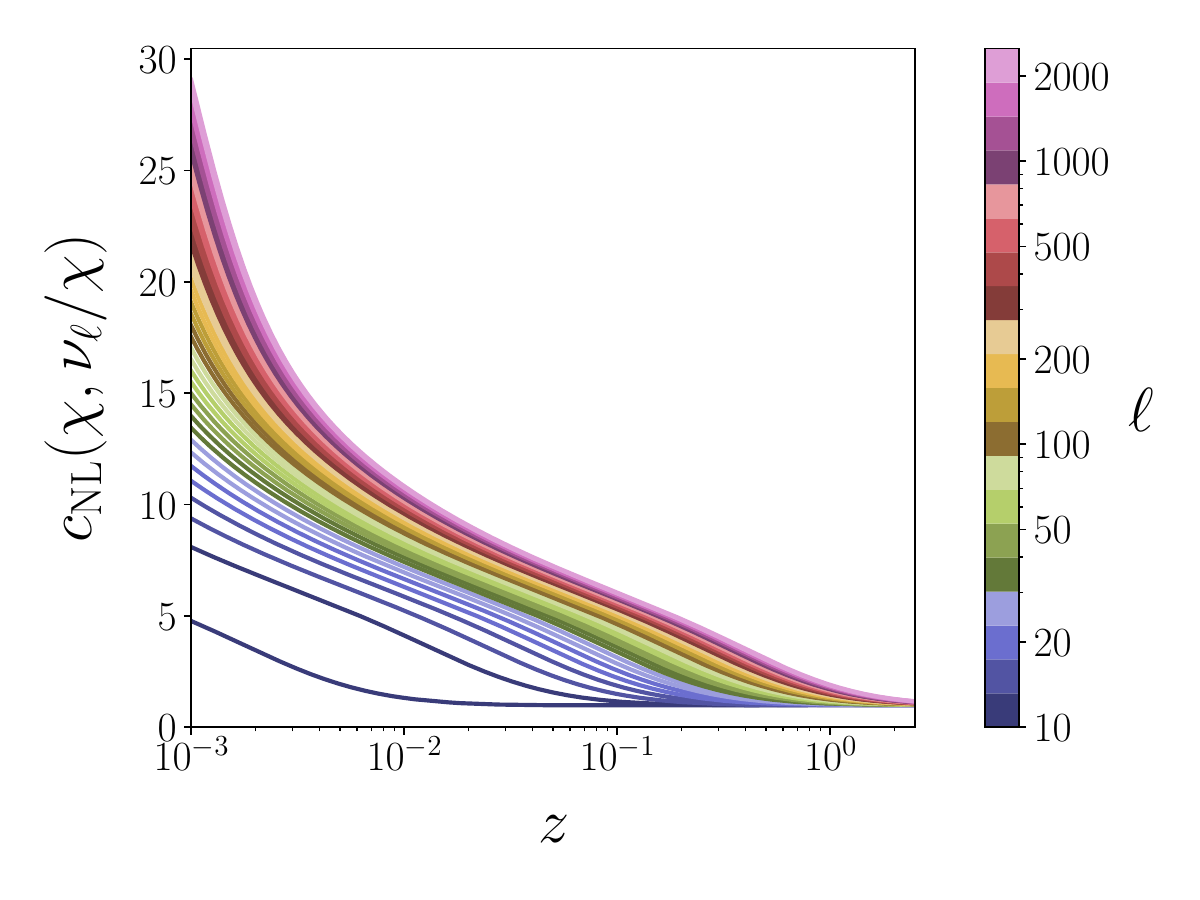}
	\caption{Non-linear correction factor calculated using HaloFit up to $z=2.5$ and $10\leq \ell\leq 2500$.}\label{fig:alpha_NL}
\end{figure}
In the Limber approximation, for a given multipole value $\ell$, the transfer functions are evaluated along the $k=\flatfrac{\nu_\ell}{\chi}$ curve in the $k$-$\chi$ plane with $\nu_\ell \equiv \ell+\flatfrac{1}{2}$. Plots of $c_{\text{NL}}(z,\nu_\ell/\chi)$ for several values of $\ell$ are shown in \autoref{fig:alpha_NL}. As would be expected, non-linear effects are most prominent at low redshift and small angular scales. Nevertheless, the integrated effect becomes noticeable even around $\ell\gtrsim 100$.

    \subsection{Euclid tomographic source distributions}	
		
Physical lensing observables are usually the average of some quantity measured along the line of sight. For example, the weak-lensing shear at some point in the sky can be estimated by determining the average ellipticity of galaxies in the bin/set of pixels corresponding to that particular direction. Theoretical calculations and predictions therefore require knowledge of {galaxy source distributions}. These are survey-specific and are usually measured in terms of redshift.

The normalised source distribution $\N(\chi)$ is defined such that $\dd P=\N(\chi)\dd\chi$ represents the probability of finding a source in the interval $[\chi,\chi+\dd\chi]$. However, these distributions are usually modelled as functions of redshift $z$. The {source redshift distribution} $n(z)$ allows us to calculate the probability of finding a source $\dd P= n(z)\dd z$ over the redshift interval $[z,z+\dd z]$. These two distributions are clearly related through
\begin{equation}
	\N(\chi) = n(z(\chi))\dv{z}{\chi} = n(z(\chi)) \frac{a_0}{a(\chi)}\H(\chi),
\end{equation}
where the final equality follows from the redshift relation in an FLRW spacetime $1+z=a_0/a$. Suppose that the survey in question is only sensitive up to some redshift $z_S$ corresponding to a conformal light-cone distance $\chi_S$. The effective, expected value of a quantity $Q$ up to this cut-off is then
\begin{equation}
	Q(\chi_S) = \int_{0}^{\chi_S}\dd\chi\, \N(\chi)\, Q(\chi) = \int_{0}^{z_S}\dd z\, n(z)\, Q(z) =   \int_{0}^{\chi_S}\dd\chi\, n(z(\chi))\, \frac{a_0}{a(\chi)}\H(\chi)\, Q(\chi).
\end{equation}
This is essentially a projected version of $Q$ along a particular line of sight. The source distributions must therefore be normalised such that
\begin{equation}
 	\int_{0}^{\chi_S}\dd\chi \,\N(\chi) = \int_{0}^{z_S}\dd z\, n(z) = 1.
 \end{equation}

\begin{figure}[htb!]
	\centering
	\includegraphics[scale=0.6]{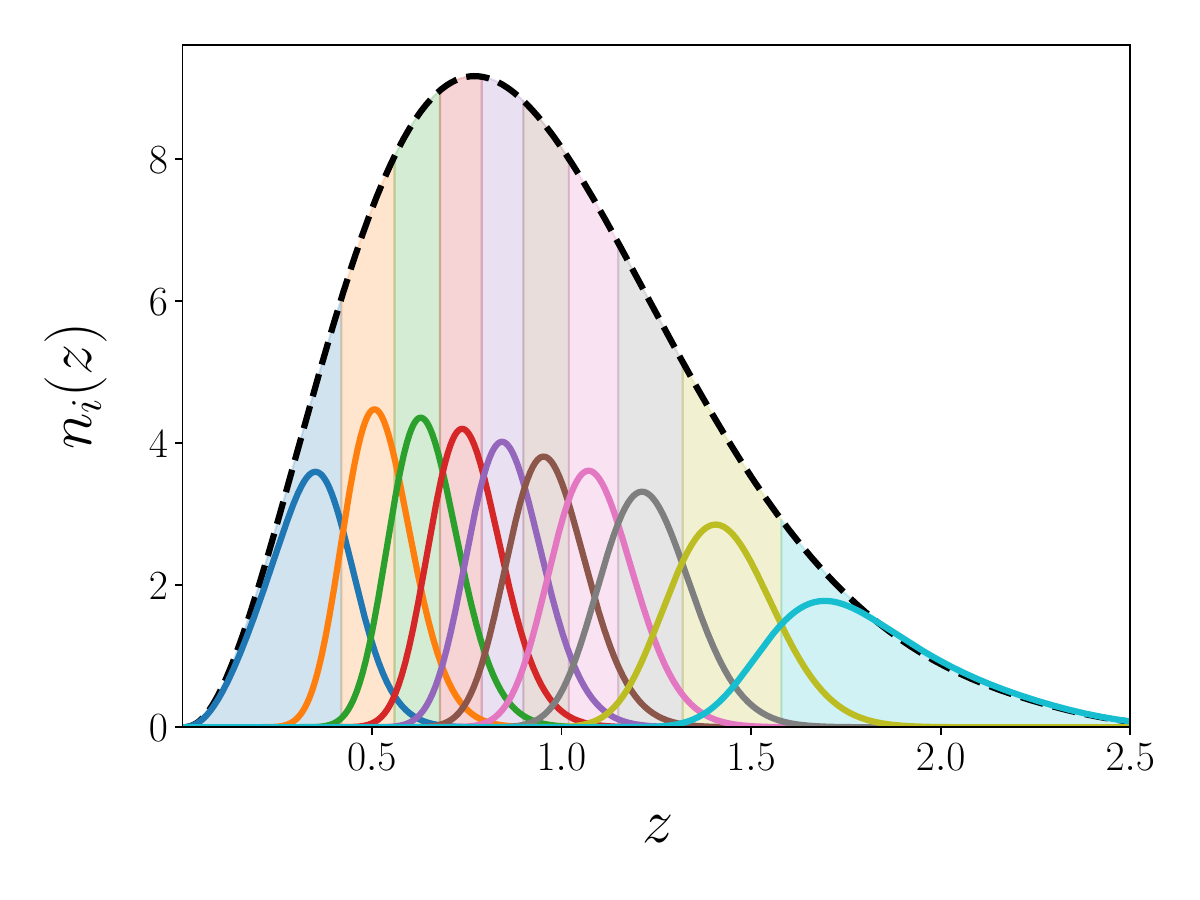}
	\caption{Plot illustrating how the full source distribution $n_\text{F}$ (dashed line) has been decomposed into 10 tomographic source distributions $n_i$ (coloured lines). The full source distribution $n_\text{F}$ has been scaled by a factor of 10 so that the area beneath it is equal to the sum of the areas beneath the tomographic distributions. Since the bins are equi-populated, the area underneath each coloured line is the same as the area of the corresponding shaded region.}
 \label{fig:Euclid_Tomographic_Distributions}
\end{figure}

Along with providing some information about time evolution, sub-dividing a population of source galaxies into redshift bins increases the amount of information available and can improve parameter constraints \cite{hu_power_1999}. In its modelling and forecasts \cite{blanchard_euclid_2020, deshpande_euclid_2024}, Euclid uses ten equi-populated bins with edges $z_i^{\pm} \in$ \{0.0010, 0.42, 0.56, 0.68, 0.79, 0.90, 1.02, 1.15, 1.32, 1.58, 2.50\}, where $1\leq i\leq 10$ is the bin label and $z^+_i = z^-_{i+1}$. The $i$th tomographic source distribution $n_i(z)$ is constructed by weighting the underlying source distribution $n(z)$ by the probability that a galaxy at redshift $z$ has a measured photometric redshift within the range $z_i^-\leq z_\text{ph}\leq z_i^+$. The resulting tomographic source distributions are shown alongside the full distribution in  \autoref{fig:Euclid_Tomographic_Distributions}. Specific details on the $n_i$ functions can be found in Appendix~\ref{sec:sourcedistrib}.

    \subsection{\boldmath \texorpdfstring{$E$}{E}- and \texorpdfstring{$B$}{B}-mode auto-correlations}\label{sec:E_B_Auto_Corr}    
The correlation between {\II} $E$-mode multipoles in \autoref{eqn:E_lm_11} calculated for tomographic bins $i$ and $j$ includes a combination of {\OI} and {\II} terms
\begin{equation}\label{eqn:E^11E^11_Correlation_Expansion}
    \begin{split}
        \expval{E^{i{\II}}_{\ell m} E^{j{\II} *}_{\ell' m'}} = \hphantom{+}& 
        \expval{\delta E^{i{\OI}}_{\ell m} \delta E^{j{\OI} *}_{\ell' m'}} + 
        \expval{\delta E^{i{\OI}}_{\ell m} \delta E^{j{\II} *}_{\ell' m'}}\\ +& 
        \expval{\delta E^{i{\II}}_{\ell m} \delta E^{j{\OI} *}_{\ell' m'}} + 
        \expval{\delta E^{i{\II}}_{\ell m} \delta E^{j{\II} *}_{\ell' m'}}.
    \end{split}    
\end{equation}
Note that corrections at order {\IO} are fully deterministic and so do not contribute to any two-point correlation functions up to order {\II}: $\expval{X^{{\IO}}Z^{{\OI}}} =X^{{\IO}}\expval{Z^{{\OI}}}  = 0$ and $\expval{X^{{\IO}}Z^{{\II}}} =X^{{\IO}}\expval{Z^{{\II}}}  = 0$ since linear stochastic perturbations have zero mean. Furthermore, since we are considering late-time anisotropic expansion, we assume that the primordial curvature perturbation $\mathcal{R}$ is a Gaussian random variable and its variance is diagonal and rotationally invariant
\begin{equation}
    \expval{\mathcal{R}(\vb{k})\mathcal{R}(\vb{k}')^{*} } = P(k)\delta(\vb{k}-\vb{k}').
\end{equation}
The primordial power spectrum $P$ wholly specifies the statistics of $\mathcal{R}$.

The first term in \autoref{eqn:E^11E^11_Correlation_Expansion} is the usual $E$-mode correlation generated by scalar perturbations and scales as $\sim \varphi^2$. Due to statistical isotropy and homogeneity at this level, this correlation function reduces to the $E$-mode power spectrum $C^{E^{i}E^{j}}_{\ell}$. Applying the Limber approximation to \autoref{eqn:C^EE_No_Limber} leads to the well-known expression
\begin{equation}\label{eqn:C^EE_Limber}
    C^{E^{i}E^{j}}_{\ell} = \frac{1}{4}\frac{(\ell+2)!}{(\ell-2)!}\int_{0}^{\chi_S}\frac{\dd \chi}{\chi^2} \, q^{i}(\chi,\chi_S)\,q^{j}(\chi,\chi_S)P\left(\frac{\nu_\ell}{\chi}\right) \left|T_\varphi\left(\chi,\frac{\nu_\ell}{\chi}\right)\right|^{2},
\end{equation}
with $\nu_\ell \equiv \ell + \flatfrac{1}{2}$.

Although the $B$-mode correlation function in \autoref{eqn:B^11B^11_Correlation} has off-diagonal contributions, in order to compare its magnitude to that of $C^{E^{i}E^{j}}_{\ell}$ we compute the $B$-mode power spectrum defined as
\begin{equation}
    C^{B^{i}B^{j}}_{\ell} \equiv (-1)^{\ell}\frac{^{B^iB^j}\!\A^{00}_{\ell \ell}}{\sqrt{2\ell+1}} = \frac{1}{2\ell+1}\sum_{m} \expval{B^{i{\II}}_{\ell m} B^{j{\II} *}_{\ell m}}.
\end{equation}
Using the Limber approximation leads to 
\begin{equation}
    \begin{split}
        C^{B^{i}B^{j}}_{\ell} = \frac{\pi}{150} &\largesum_{I = \pm 1} \frac{(\ell+I+2)!}{(\ell+I-2)!}\frac{{\left(^2F_{\ell 2 \ell+I}\right)^{2}}}{2\ell+1}\\
        &\cdot\int_{0}^{\chi_S}\,\frac{\dd \chi}{\chi^2} q^{i}(\chi,\chi_S)q^{j}(\chi,\chi_S)P\left(\frac{\nu_\ell+I}{\chi}\right) \left|\alpha(\chi)T_\varphi\left(\chi,\frac{\nu_\ell+I}{\chi}\right)\right|^{2},    
    \end{split}    
\end{equation}
with $|\alpha|^2\equiv \alpha_{ij}\alpha^{ij}$.

The intrinsic ellipticities of galaxies introduce unavoidable uncertainty into weak lensing observations called shape noise. Although real-world lensing analyses use far more sophisticated methods of quantifying this uncertainty contribution, a crude estimate can be obtained by modelling it as white noise that is un-correlated between different redshift bins. For both the $E$- and $B$-mode power spectra, this approximation of the statistical error amounts to
\begin{equation}
    \Delta C^{ij}_{\ell} = \bigg[{\frac{2}{(2\ell+1)f_{\text{sky}}}}\bigg]^{1/2}\,\frac{\expval{\gamma_{\text{int.}}^2}}{\bar{N}_{i}}\delta_{ij},
\end{equation}
where $f_{\text{sky}} = 0.35$ is the fraction of the sky covered by the Euclid survey, $\expval{\gamma_{\text{int.}}^2} = 0.3^2$ is the intrinsic ellipticity variance, and  $\bar{N}_{i}\approx 2.7$ arcmin$^{-2}$ is the number density of galaxies in the selected redshift bin. Euclid is expected to measure around 1.5 billion galaxy shapes in its survey area corresponding to a total density of $\bar{N}\approx 27$ arcmin$^{-2}$, which is then split among its ten equi-populated bins. 

%\note{*[How do we quantify error between different bins in a simple way? One could maybe use some kind of IA model, but is this overkill?]*}

\begin{figure}[htb!]
	\centering
	\includegraphics[scale=0.6]{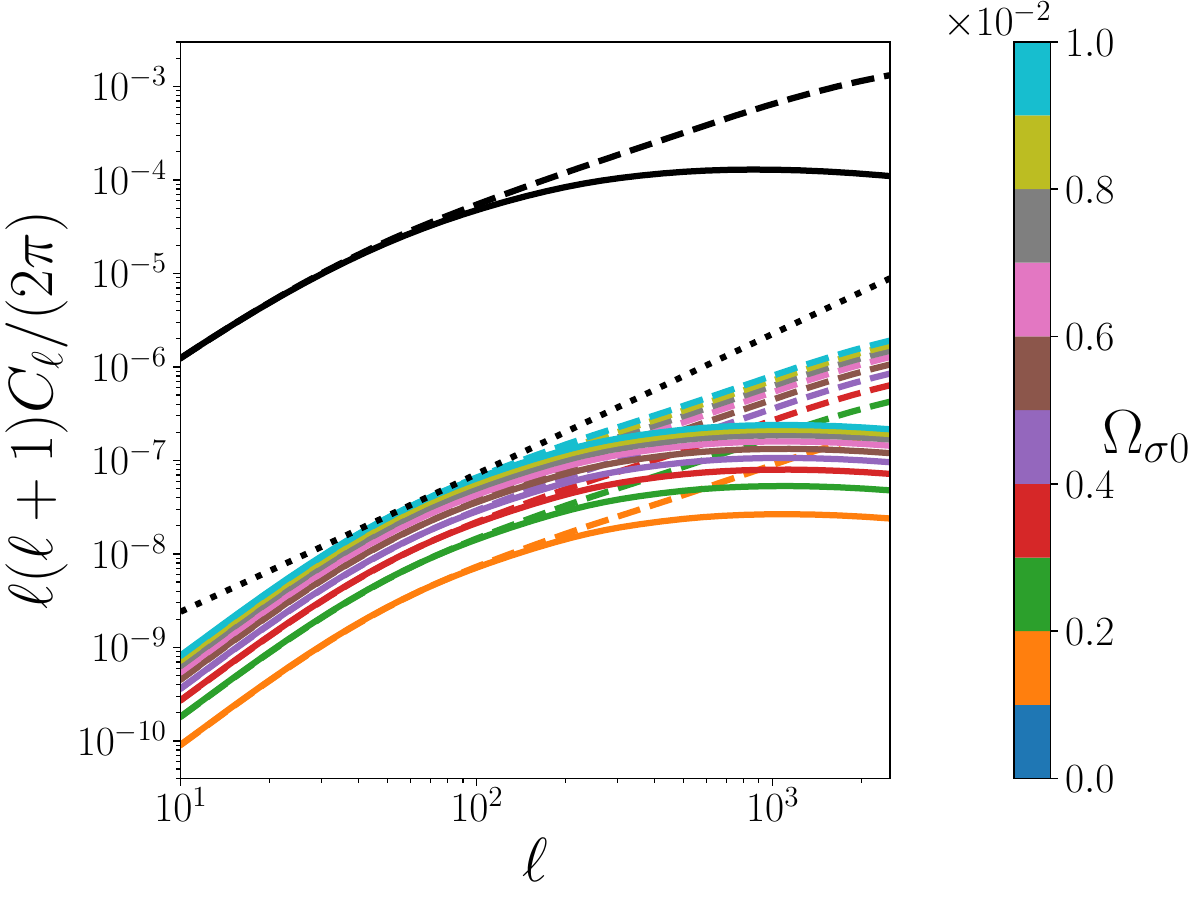}
	\caption{Linear (solid) and HaloFit (dashed) $E$-mode (black) $B$-mode (multi-coloured) power spectra for full source distribution $n_\text{F}$ plotted alongside shape noise (dotted) over the range $10 \leq \ell \leq 2500$.}\label{fig:C^BB_ell_Varying_Sigma}
\end{figure}

The behaviour of $C^{BB}_\ell$, which scales as $\sim \varphi^2 (\sigma/\H)^2$,  is shown for several values of $\Omega_{\sigma 0}$ in  \autoref{fig:C^BB_ell_Varying_Sigma}, alongside the dominant $E$-mode power spectrum contribution \autoref{eqn:C^EE_Limber}. These power spectra were computed using the full underlying source distribution $n_\text{F}$ and hence the corresponding shape noise is reduced by a factor of 10 compared to a single tomographic bin. Although dialling $\Omega_{\sigma0}$ up to a value of $10^{-2}$ (i.e. $\sigma_0/\H_0 \approx 10^{-1}$) naturally increases $C^{BB}_\ell$, it never exceeds the shape noise baseline even in this best-case scenario. This suggests that any {\II}$\times${\II} corrections to angular power spectra will likely be very difficult, and maybe even impossible, to detect with upcoming surveys. A similar issue was found by \cite{pitrou_weak_2013} in their letter \cite{pereira_weak-lensing_2016} when trying to obtain a rough estimate of $\sigma_0/\H_0$ using the $B$-mode auto-correlation data from CFHTLenS \cite{kitching_3d_2014}. It is for this reason that we suggest looking at the $E$-$B$ cross-correlation, since its leading-order contribution is {\OI}$\times${\II} and hence scales like $\sim \varphi^2 (\sigma/\H)$. Moreover, to a reasonable approximation, shape noise should not correlate between $E$- and $B$-modes due to their parity properties. Making use of tomography, along with providing information about anisotropy evolution, will help to reduce statistical error since shape noise does not correlate over large redshift separations.  

    \subsection{\boldmath \texorpdfstring{$E$}{E}-\texorpdfstring{$B$}{B} cross-correlation}
Unlike in \autoref{eqn:E^11E^11_Correlation_Expansion}, the cross-correlation between the $E$ and $B$ multipoles only has two terms
\begin{equation}
    \expval{E^{i{\II}}_{\ell m} B^{j{\II} *}_{\ell' m'}} =  
        \expval{\delta E^{i{\OI}}_{\ell m} \delta B^{j{\II} *}_{\ell' m'}} + 
        \expval{\delta E^{i{\II}}_{\ell m} \delta B^{j{\II} *}_{\ell' m'}}\,.
\end{equation}
Keeping only the dominant first term leads to the BipoSH coefficient
\begin{equation}\label{eqn:EB_BipoSH_Coeff_Dominant}
    ^{E^iB^j}\!\mathcal{A}^{LM}_{\ell \ell\pm 1} = -\frac{i}{4}\frac{{^2F_{\ell \pm 1, 2, \ell}}}{\sqrt{5}}\frac{(\ell+2)!}{(\ell-2)!}\,\mathcal{P}_{\ell M}^{ij}\delta^{2L}\,,
\end{equation}
which is pure imaginary and non-zero only for $L=2$. The Limber-approximated quantity\footnote{Note that our definition differs from the one in \cite{pitrou_weak-lensing_2015,pereira_weak-lensing_2016} through the index $M\longmapsto -M$.} 
\begin{equation}\label{eqn:P_lM_defn}
	\mathcal{P}_{\ell M}^{ij} \equiv \int_{0}^{\chi_S} \frac{\dd\chi}{\chi^{2}}\, q^{i}(\chi,\chi_S)\,q^{j}(\chi,\chi_S) P\left(\frac{\nu_\ell}{\chi}\right)\alpha_{2,(-M)}^{{\IO}}(\chi)\left|T_{\varphi}\left(\chi,\frac{\nu_\ell}{\chi}\right)\right|^{2}\,,
\end{equation}
contains the five degrees of freedom present in the metric shear $\bs{\sigma}$. Note that $\mathcal{P}_{\ell M}^{ij}$ is only symmetric under the interchange of $i$ and $j$ in the Limber regime.

\begin{figure}[htb]
	\centering
	\includegraphics[scale=0.5]{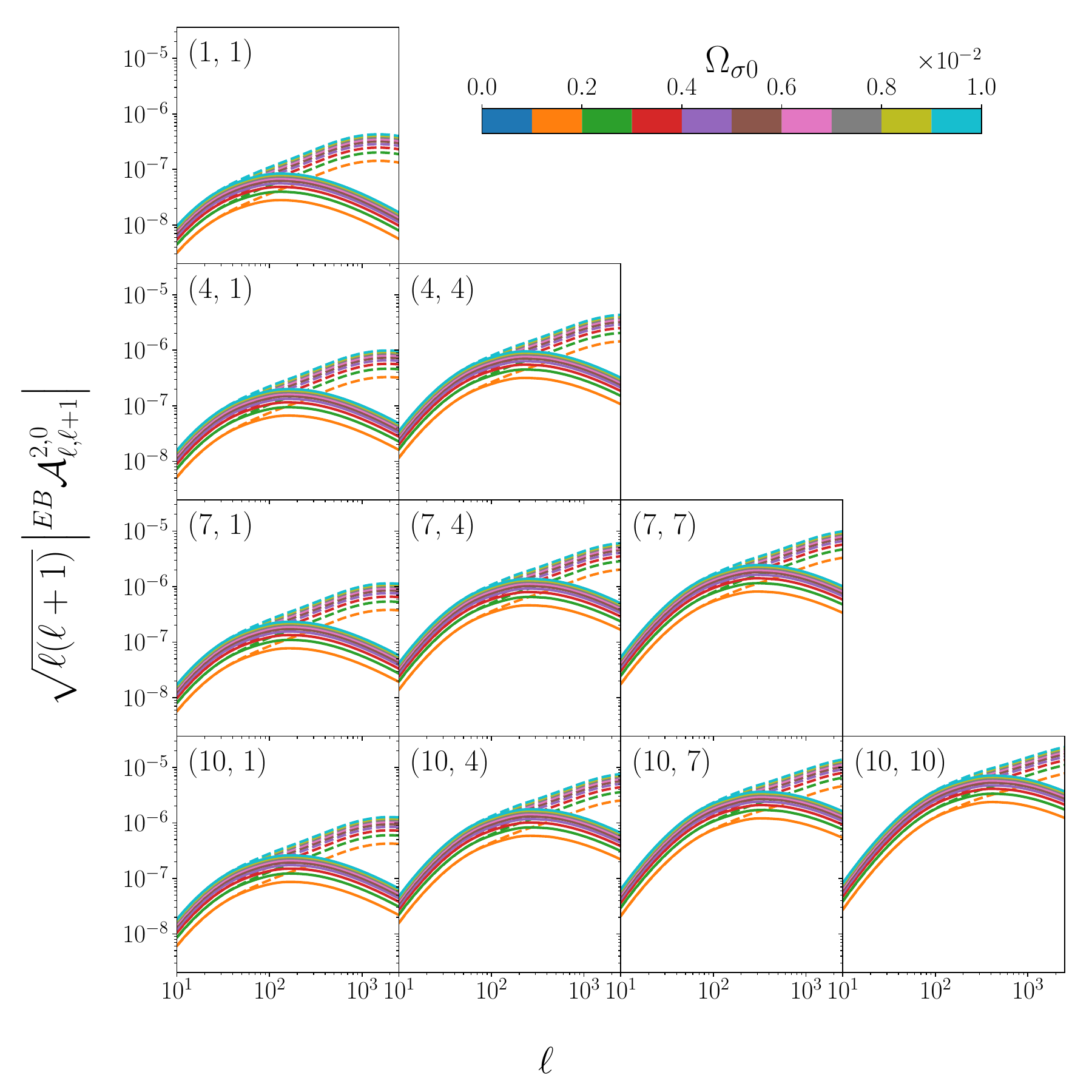}
	\caption{Linear (solid) and HaloFit (dashed) tomographic {\biposh} coefficient ${^{E^iB^j}\!\mathcal{A}^{2,0}_{\ell, \ell + 1}}$ for $10 \leq \ell \leq 2500$ calculated using ten values of $\Omega_{\sigma 0}\leq 10^{-2}$. The auto- and cross-correlations between redshift bins 1, 4, 7, and 10 are shown. The  scaling $\sqrt{\ell(\ell+1)}$ is used to highlight the effect of varying $\Omega_{\sigma 0}$ and the differences between the linear and HaloFit spectra. the $\ell-1$ coefficient ${^{E^iB^j}\!\mathcal{A}^{2,0}_{\ell, \ell - 1}}$ is not shown because it is visually indistinguishable from the $\ell+1$ case.} 
 \label{fig:Al_EB}
\end{figure}

As can be seen from \autoref{eqn:Beta_2m_Multipoles}, imposing axisymmetry about the $z$-axis forces the {\biposh} coefficients in \autoref{eqn:EB_BipoSH_Coeff_Dominant} to be non-zero only for $M=0$.  \autoref{fig:Al_EB} depicts the coefficient ${^{E^iB^j}\!\mathcal{A}^{2,0}_{\ell, \ell + 1}}$ in the case of axisymmetry. As one would expect, the signal is markedly stronger at higher redshifts while non-linear corrections are more noticeable at lower redshifts. As with the auto-correlation power spectra in  \autoref{fig:C^BB_ell_Varying_Sigma}, the linear and HaloFit coefficients begin to deviate significantly from one another around $\ell\sim 100$. This observation will help us to identify a cut-off scale for our analysis in \S \ref{sec:Estimator_SNR}.

%%%%%%%%%%%%%%%%%%%%%%%%%%%%%%%%%%%%%%%%%%%%%%%%%%%%%%%%%%%%%%%%%%%%%%%%%%%%%%%%%%%%%%%%%%%%%%%%%%%%%%%%%

\section{\boldmath Detectability of the \texorpdfstring{$E$}{E}-\texorpdfstring{$B$}{B} cross-correlation}\label{sec:Estimator_SNR}

As we mentioned in \S \ref{sec:Angular_Spectra}, cosmic variance prohibits one from measuring each {\biposh} coefficient from a single sky map. Instead, one needs to combine a subset of these coefficients in order to isolate some theoretically-predicted quantity. In this section, we construct a simple estimator for the quantity ${\P}_{\ell M}^{ij}$ in \autoref{eqn:P_lM_defn} and investigate its expected signal-to-noise ratios for a Euclid-like survey.

    \subsection{Estimator}
Suppose we have a set of $E$- and $B$-mode multipoles obtained from a map, which we choose to represent as $\hat{E}^i_{\ell m}$ and $\hat{B}^i_{\ell m}$, respectively. An unbiased statistical estimator for the $E$-$B$ {\biposh} coefficient is then 
\begin{equation}\label{eqn:EB_BipoSH_Estimator}
    ^{E^iB^j}\!\hat{\A}^{LM}_{\ell \ell'} = \sqrt{2L+1} \largesum_{m,m'}(-1)^{L+m} \mqty(\ell & \ell' & L \\ -m & m' & M) \hat{E}^i_{\ell m}\hat{B}^{j*}_{\ell' m'}\,.
\end{equation}
In order to simplify our analysis and to quantify deviations from isotropy, we adopt a `null hypothesis' of statistical isotropy. Under this assumption, $\hat{B}^i_{\ell m}$ coefficients are composed solely of shape noise while the $\hat{E}^i_{\ell m}$ contain shape noise as well as the {\OI} $E$-mode signal \autoref{eqn:E_01_Multipoles}. If we additionally assume that the $E$- and $B$-mode multipoles obey Gaussian statistics, we can analytically compute the isotropic part of the covariance of the estimator \autoref{eqn:EB_BipoSH_Estimator} as
\begin{equation}\label{eqn:EB_BipoSH_Estimator_Covariance} 
    \text{Cov}\left({^{E^iB^j}\!\hat{\A}^{L_1M_1}_{\ell_1 \ell_1'}},{^{E^kB^l}\!\hat{\A}^{L_2M_2}_{\ell_2 \ell_2'}}\right)_\text{SI} = \frac{1}{f_{\text{sky}}} {\left(C^{E^i E^k}_{\ell_1}\right)_{\text{SI}}} {\left(C^{B^j B^l}_{\ell_1'}\right)_{\text{SI}}}\delta_{\ell_1\ell_2}\delta_{\ell_1'\ell_2'} \delta^{L_1 L_2}\delta^{M_1M_2}\,,
\end{equation}
where the statistically-isotropic angular power spectra are defined through
\begin{subequations}
    \begin{align}
        \expval{\hat{E}^i_{\ell m} \hat{E}^{j*}_{\ell' m'}}_{\text{SI}} &\equiv {\left(C^{E^i E^j}_{\ell}\right)_{\text{SI}}}\delta_{\ell \ell'}\delta_{m m'} = \left(\frac{\expval{\gamma_{\text{int.}}^2}}{\bar{N}_{i}}\delta_{ij} + C^{E^iE^j}_\ell\right)\delta_{\ell \ell'}\delta_{m m'}\,, \\
        \expval{\hat{B}^i_{\ell m} \hat{B}^{j*}_{\ell' m'}}_{\text{SI}} &\equiv {\left(C^{B^i B^j}_{\ell}\right)_{\text{SI}}}\delta_{\ell \ell'}\delta_{m m'} = \frac{\expval{\gamma_{\text{int.}}^2}}{\bar{N}_{i}}\delta_{ij}\delta_{\ell \ell'}\delta_{m m'}\,, \\        
        \expval{\hat{E}^i_{\ell m} \hat{B}^{j*}_{\ell' m'}}_{\text{SI}} &\equiv {\left(C^{E^i B^j}_{\ell}\right)_{\text{SI}}}\delta_{\ell \ell'}\delta_{m m'} = 0\,,
    \end{align}
\end{subequations}
and we have divided through by the observed sky fraction $f_{\text{sky}}$ in order to account for the increase in variance caused by a partially-surveyed sky. 

By inverting \autoref{eqn:EB_BipoSH_Coeff_Dominant}, we can construct the {simple} estimator 
\begin{equation}\label{eqn:P_lM_Estimator}
    \hat{\mathcal{P}}_{\ell M}^{ij} = 2\sqrt{5}\,i\,\frac{(\ell-2)!}{(\ell+2)!}\largesum_{I=\pm 1}\frac{^{E^iB^j}\!\hat{\A}^{2M}_{\ell, \ell+I}}{{^2F_{\ell+I, 2, \ell}}}\,,
\end{equation}
which weights both the $\ell+1$ and $\ell-1$ contributions equally. Using \autoref{eqn:EB_BipoSH_Estimator_Covariance}, we find that the isotropic part of this estimator's covariance is given by
\begin{equation}\label{eqn:P_lM_Estimator_Covariance}
    \text{Cov}\left(\hat{\mathcal{P}}_{\ell M}^{ij}, \hat{\mathcal{P}}_{\ell' M'}^{kl}\right)_{\text{SI}} = \frac{20}{f_{\text{sky}}} \left[ \frac{(\ell-2)!}{(\ell+2)!} \right]^2 \largesum_{I=\pm 1}\frac{{\left(C^{E^i E^k}_{\ell}\right)_{\text{SI}}} {\left(C^{B^j B^l}_{\ell+I}\right)_{\text{SI}}}}{\left({^2F_{\ell+I, 2 ,\ell}}\right)^2} \delta_{\ell \ell'}\delta_{MM'}.
\end{equation}

One could also construct an estimator for $\mathcal{P}_{\ell M}^{ij}$ using the {\biposh} coefficients $^{E^iE^j}\!\A^{2M}_{\ell, \ell}$ and $^{E^iE^j}\!\A^{2M}_{\ell, \ell \pm 2}$. However, if we consider the variance of the corresponding estimator, 
\begin{equation}   \text{Var}\left(^{E^iE^j}\!\hat{\A}^{2M}_{\ell\ell'}\right)_{\text{SI}} = \frac{1}{f_{\text{sky}}}\left[ {\left(C^{E^i E^i}_{\ell}\right)_{\text{SI}}} {\left(C^{E^j E^j}_{\ell'}\right)_{\text{SI}}} + {\left(C^{E^i E^j}_{\ell}\right)_{\text{SI}}^2} \delta_{\ell\ell'}\right],
\end{equation}
we see that it contains extra contributions from shape noise and the $E$-mode power spectrum. An estimator constructed from these coefficients will therefore have lower signal-to-noise than the one in \autoref{eqn:P_lM_Estimator}. This is another reason why we believe the cross-correlation of the shear $E$-modes and $B$-modes is a better way to detect large-scale anisotropy than the $E$-$E$ and $B$-$B$ auto-correlations.

    \subsection{Signal-to-noise ratios}

\begin{figure}[htb]
	\centering
	\includegraphics[width=\textwidth]{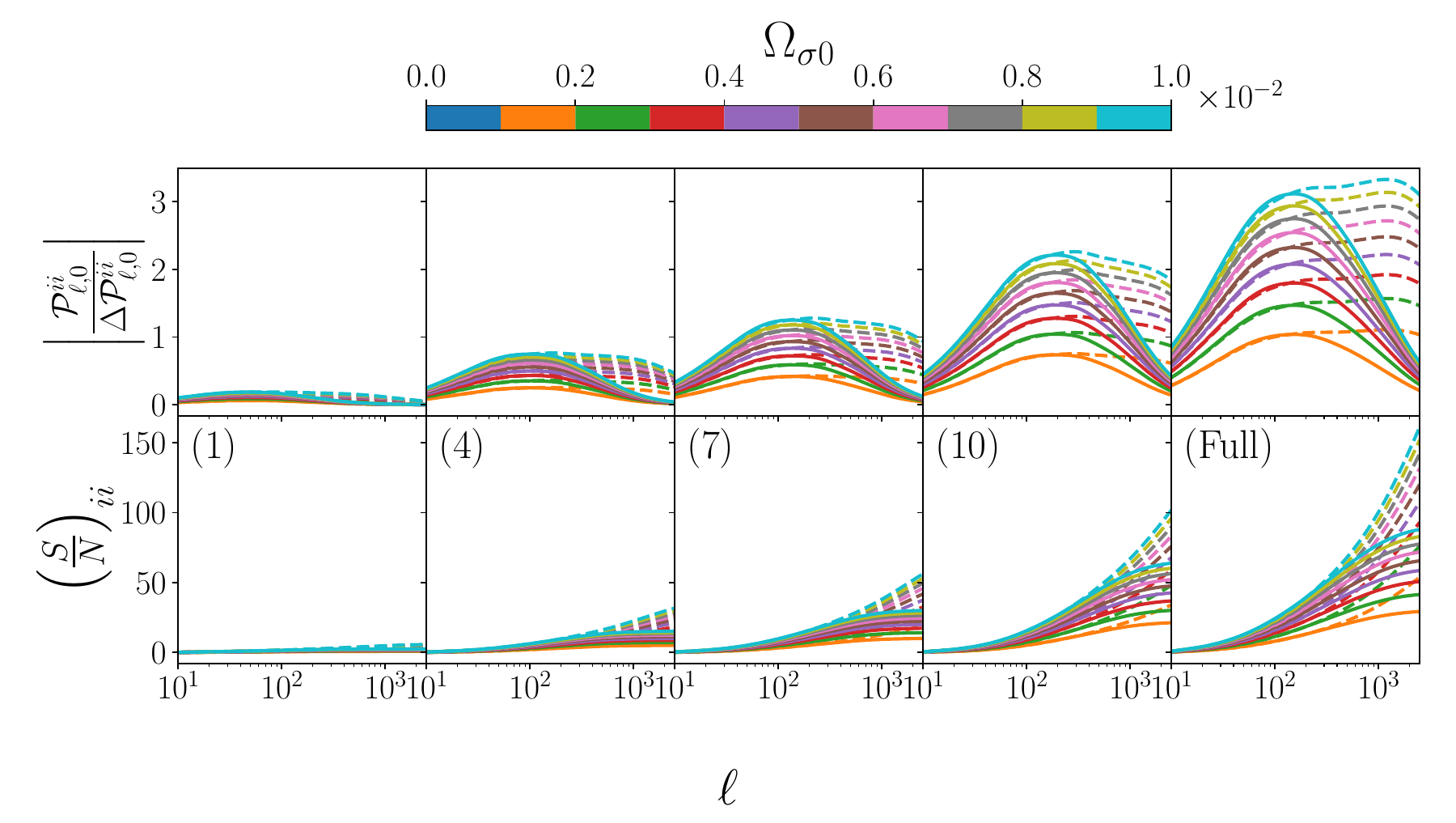}
	\caption{Individual (top) and cumulative (bottom) signal-to-noise ratios for $\hat{\P}^{ii}_{\ell,0}$ computed for redshift bins $i=1, 4, 7, 10$, as well as the full redshift range. The linear (solid) and HaloFit (dashed) ratios have been calculated for ten values of $\Omega_{\sigma 0}\leq 10^{-2}$. Note that we use $\ell_{\text{min}}=10$ for the cumulative SNR.}
 \label{fig:PlM_SNR_Diag_lmax}
\end{figure}

In principle, one could compute the total signal-to-noise ratio (SNR)  
\begin{equation}
    \left(\frac{S}{N}\right)^2_\text{tot} \equiv \largesum_{\substack{\ell,\ell'\\ M,M'}} \largesum_{i,j,k,l} {\mathcal{P}}_{\ell M}^{ij} \text{Cov}^{-1}\left(\hat{\mathcal{P}}_{\ell M}^{ij}, \hat{\mathcal{P}}_{\ell' M'}^{kl}\right)_{\text{SI}} {\mathcal{P}}_{\ell' M'}^{kl~*},
\end{equation}
which combines all multipole and tomographic information using the inverse of the covariance matrix in \autoref{eqn:P_lM_Estimator_Covariance}. However, in order to investigate how the potential strength of the signal varies with redshift, we consider each tomographic correlation separately and calculate the cumulative SNR for each $\hat{\mathcal{P}}_{\ell M}^{ij}$ as 
\begin{equation}
    \left(\frac{S}{N}\right)^2_{ij} = \largesum_{\ell = \ell_{\text{min}}}^{\ell_{\text{max}}} \left(\frac{\mathcal{P}_{\ell M}^{ij}}{\Delta \mathcal{P}^{ij}_{\ell M}}\right)^2 \quad \mbox{where} \quad
    \Delta \mathcal{P}^{ij}_{\ell M}\equiv \bigg[{\text{Var}\left(\hat{\mathcal{P}}_{\ell M}^{ij}\right)_{\text{SI}}}\bigg]^{1/2}\,.
\end{equation}
%where the deviation $\Delta \mathcal{P}^{ij}_{\ell M}\equiv \Big[{\text{Var}\left(\hat{\mathcal{P}}_{\ell M}^{ij}\right)_{\text{SI}}}\Big]^{1/2}$ is simply the square root of the variance of each individual $\hat{\mathcal{P}}_{\ell M}^{ij}$. 
Heuristically, the cumulative SNR is a measure of the constraining power of a particular observable and is closely related to Fisher information. In order to simplify our analysis and argument, we consider only the diagonal tomographic correlations and set $i=j$ from now on.

The individual and cumulative SNRs for the diagonal (i.e. $i=j$) tomographic estimators $\hat{\P}^{ii}_{\ell,0}$ are shown in  \autoref{fig:PlM_SNR_Diag_lmax}. As can be seen from the top row of this figure, cosmic variance strongly suppresses the SNRs for $\ell \lesssim 10$. We also note that the Limber approximation that we have used breaks down on these largest scales. Moreover, like with the {\biposh} coefficients shown in  \autoref{fig:Al_EB}, the linear and non-linear SNRs in  \autoref{fig:PlM_SNR_Diag_lmax} begin to deviate significantly for $\ell \gtrsim 100$. Nevertheless, it is evident that much of the constraining power derives from larger scales corresponding to $10\lesssim \ell \lesssim 100$. With this in mind, we choose to set $\ell_{\text{min}}=10$ and $\ell_{\text{max}}=100$ for the remaining analysis. These conservative bounds ensure that we do not miss too much information on the largest scales while avoiding any additional $B$-mode sources on smaller scales.

\begin{figure}[htb]
	\centering
	\includegraphics[scale=0.6]{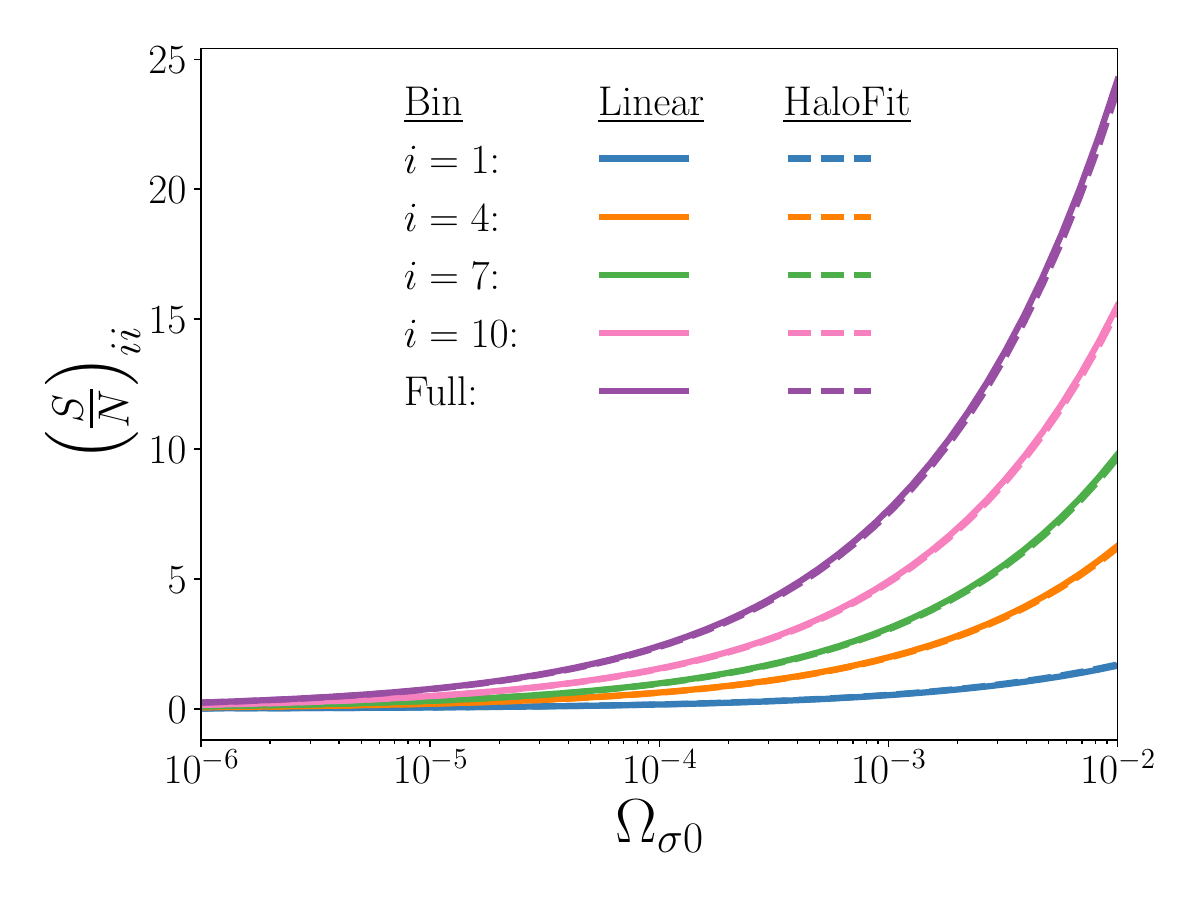}
	\caption{Linear (solid) and HaloFit (dashed) cumulative $\hat{\P}^{ii}_{\ell,0}$ signal-to-noise ratios summed from $\ell_{\text{min}}=10$ up to $\ell_{\text{max}} = 100$ as a function of $\Omega_{\sigma 0}$. The different colours are used to represent the redshift bins $i=1, 4, 7, 10$, as well as the full redshift range.}
 \label{fig:PlM_SNR_Diag_lmax_var_Omega}
\end{figure}

In \autoref{fig:PlM_SNR_Diag_lmax_var_Omega}, we see that our choice of $\ell_{\text{max}}=100$ ensures that the contribution of any non-linear effects is sufficiently small. Naturally, the cumulative SNR increases in tandem with the strength of $\Omega_{\sigma 0}$. Again, we see that the majority of the constraining power is provided by the higher redshift bins. Nevertheless, it is clear that combining all tomographic information will yield a higher SNR and tighter constraints than using the (non-tomographic) full distribution by itself.

We highlight, however, that this somewhat rudimentary analysis can certainly be improved and expanded upon. For example, more sophisticated versions of the estimators \autoref{eqn:EB_BipoSH_Estimator} and \autoref{eqn:P_lM_Estimator} which make use of inverse-variance weighting and band-power averaging would certainly improve the quality of the observed signal. We also acknowledge that our analysis does not fully account for the effects of a masked sky -- which can lead to spurious correlations between different scales. That being said, we believe that our SNR estimates are likely an underestimate of what is possible with surveys like Euclid. Most importantly, we emphasise the cross-correlation of the $E$- and $B$-modes of cosmic shear as a possible way to constrain late-time violation of the Cosmological Principle.

%%%%%%%%%%%%%%%%%%%%%%%%%%%%%%%%%%%%%%%%%%%%%%%%%%%%%%%%%%%%%%%%%%%%%%%%%%%%%%%%%%%%%%%%%%

\section{Concluding remarks}
This work revisits and builds upon the analysis of \cite{pitrou_weak-lensing_2015,pereira_weak-lensing_2016}. We reviewed their formalism and argument for identifying the dominant contribution to the $B$-mode shear and computed the weak-lensing signal generated by a phenomenological model of late-time anisotropic expansion with axisymmetry. We found in \S \ref{sec:E_B_Auto_Corr} that the strength of the $B$-mode power spectrum is below Euclid's detectability threshold for all realistic values of $\Omega_{\sigma0}$. We then argued that the cross-correlation of $E$- and $B$-mode shear offers a better prospect for detecting large-scale anisotropy. 

Using the apparatus of bipolar spherical harmonics, we constructed a simple estimator (\autoref{eqn:P_lM_Estimator}) for an observable which, in principle, contains all information necessary to reconstruct the magnitude and direction of anisotropic expansion. By analysing the signal-to-noise ratios for this statistical estimator, we found that much of the constraining power of the $E$-$B$ cross-correlation derives predominantly from angular scales corresponding to $\ell \lesssim 100$ (where the $B$-mode signal is not contaminated by any non-linear effects) and higher redshifts. Exploiting available tomographic information will also significantly tighten any constraints that can be made using this method.  

The primary purpose of this work is not to provide a rigorous analysis of the potential constraining power of Euclid, but to emphasise the $E$-$B$ cross-correlation as an additional test of the Cosmological Principle. We hope that our analysis will provide a general basis for future investigations into this topic.

\acknowledgments{Numerical calculations were performed using \texttt{NumPy} \cite{harris_array_2020} and \texttt{SciPy} \cite{virtanen_scipy_2020}. Figures were produced using \texttt{matplotlib} \cite{hunter_matplotlib_2007}.} 
JA and RM are supported by the South African Radio Astronomy Observatory and National Research Foundation (grant no. 75415). JL thanks J.-P. Uzan and C. Pitrou for valuable discussions at an early stage of this project.
%%%%%%%%%%%%%%%%%%%%%%%%%%%%%%%%%%%%%%%%%%%%%%%%%%%%%%%%%%%%%%%%%%%%%%%%%%%%%%%%%%%%%%%%%%

\newpage
\cleardoublepage

\appendix
    \section{Bianchi-I and the weak shear limit}
        \subsection{Perturbed metric}\label{sec:Perturbed_Metric}
The most general perturbed Bianchi-I line element takes the form
\begin{equation}\label{eqn:B_I_Perturbed_Line_Element_1}
	\dd s^{2} = a^{2}{\left[-(1+2S)\dd\eta^{2}+2{\cal V}_{i}\dd x^{i}\dd\eta+(\gamma_{ij}+{\cal T}_{ij})\dd x^{i}\dd x^{j}\right]},
\end{equation}
where {$S$} is a scalar and the remaining functions are defined as \cite{pitrou_weak-lensing_2015,pitrou_theory_2007,pitrou_predictions_2008}
\begin{subequations}
	\begin{align}
		{{\cal V}_i} &\equiv {\partial_{i}{\cal V} +V_i} \label{eqn:B^i_defn} \\
		{{\cal T}_{ij}} &\equiv	2C\left(\gamma_{ij}+\frac{\sigma_{ij}}{\H}\right)+{2\partial_{i}\partial_{j}{\cal T}+2\partial_{(i}{\cal T}_{j)}+2\mathbb{T}_{ij}}.
	\end{align}
\end{subequations}
We have therefore introduced four scalar-valued functions ({$S$, ${\cal V}$, $C$, and ${\cal T}$}), two vector-valued functions ({$V_{i}$ and ${\cal T}_{i}$}), and one tensor-valued function ({$\mathbb{T}_{ij}$}). The inclusion of $\sigma_{ij}/\H$ in the perturbed metric is not mandatory, but is justified a posteriori by the simpler transformation properties it generates \cite{pitrou_theory_2007}. The vector variables are subject to the usual transversality and conditions
\begin{equation}\label{eqn:E_B_Transverse}
	\partial_{i}{ V^i}= \partial_{i}{{\cal T}^i} = 0,
\end{equation}
while the symmetric tensor perturbation is traceless in addition to being transverse
\begin{equation}
	\partial_{i}{\mathbb{T}^{ij}} = {\mathbb{T}\indices{^i_i}}=0.
\end{equation}
Gauge freedom allows us to place a maximum of four restrictions on the perturbed degrees of freedom. Following \cite{pitrou_weak-lensing_2015}, we make use of a gauge defined by the conditions
\begin{equation} \label{eqn:B_I_Gauge_Condition}
	{{\cal V}={\cal T}={\cal T}^i}=0.
\end{equation}
The Bardeen variables then simplify to
%\begin{subequations}
%	\begin{multicols}{2}\noindent
		\begin{align}
			\Phi = {S}\,, ~~
			\Psi = -C \,,~~
		%\end{align}
		%\columnbreak		
		%\begin{align}
				\Theta_{i} = {V_i}\,,~~
				{\mathbb{T}_{ij}}\,, %&\hphantom{=}
		\end{align}
%	\end{multicols}\noindent
%\end{subequations}
and the line element \autoref{eqn:B_I_Perturbed_Line_Element_1} becomes
\begin{equation}\label{eqn:B_I_Pert_Line_Elem}
	\dd s^{2} = a^{2}\left\{-(1+2\Phi)\dd \eta^{2}+{\Theta_i}\dd x^{i}\dd\eta + \left[\gamma_{ij}-2\Psi\left(\gamma_{ij}+\frac{\sigma_{ij}}{\H}\right)+2{\mathbb{T}_{ij}}\right]\dd x^{i}\dd x^{j}\right\}.
\end{equation}
        \subsection{Procedure for extracting contributions at arbitrary order}\label{sec:Arbitrary_Pert}
The issue of obtaining contributions of arbitrary order from a general expression can be solved in a straightforward and systematic way. To achieve this, we introduce two smallness parameters $\epsilon_1,\epsilon_2\ll 1$ which control the order of our expansions. These smallness parameters are merely a bookkeeping device that we use to derive perturbed expressions and we shall absorb them into the definitions of relevant quantities after determining these expressions. We write a formal expansion of some quantity $Q$ as
\begin{equation}
	Q = Q^{\OO} + \sum_{j,k=0}^{\infty} \frac{1}{j!k!}\delta Q^{\{j,k\}}\epsilon_1^j  \epsilon_2^k, 
\end{equation}
where $Q^{\OO}$ is the (un-perturbed) background value of $Q$, $\delta Q^{\{j,k\}}$ is the correction of order $\{j,k\}$ to this quantity and we have defined $\delta Q^{\OO}\equiv 0$. A truncated series up to order $\{n,m\}$ defines the perturbed quantity of order $\{n,m\}$
\begin{equation}
 Q^{\{n,m\}} = Q^{\OO} + \sum_{j,k=0}^{n,m} \frac{1}{j!k!}\delta Q^{\{j,k\}}\epsilon_1^j  \epsilon_2^k.
\end{equation}
For clarity, note that the object $Q^{\{n,m\}}$ contains terms of the order \emph{up to and including} $\{n,m\}$ whereas $\delta Q^{\{n,m\}}$ contains \emph{only} terms of the order $\{n,m\}$. We can extract the ${\{n,m\}}$ correction to $Q$ using
\begin{equation}\label{eqn:Order_Extraction}
	\delta Q^{\{n,m\}} = \left.\frac{\partial^{n+m}}{\partial\epsilon_1^{n}\partial \epsilon_2^{m}}Q\right|_{\epsilon_1,\epsilon_2 = 0}.
\end{equation}
As an example, consider a quantity of the form $Q=PS$. The corrections up to ${\II}$ are given by
\begin{subequations}
\begin{align}
	\delta Q^{{\IO}} &= \left.\pdv{\epsilon_1}Q\right|_{\epsilon_1,\epsilon_2 = 0} = \delta P^{{\IO}}S^{\OO} + P^{\OO} \delta S^{{\IO}} \\ 	
		\delta Q^{{\OI}} &= \left.\pdv{\epsilon_1}Q\right|_{\epsilon_1,\epsilon_2 = 0} = \delta P^{{\OI}}S^{\OO} + P^{\OO}\delta S^{{\OI}}\\
		\begin{split}\delta Q^{\{1,1\}} &= \left.\frac{\partial^{2}}{\partial\epsilon_1\partial \epsilon_2}Q\right|_{\epsilon_1,\epsilon_2 = 0} = \delta P^{{\II}}S^{\OO}		 + \delta P^{{\IO}} \delta S^{{\OI}}\\ &\hphantom{\delta Q^{\{1,1\}} = \left.\frac{\partial^{2}}{\partial\epsilon_1\partial \epsilon_2}Q\right|} + \delta P^{{\OI}} \delta S^{{\IO}}+P^{\OO} \delta S^{{\II}}.\end{split}
\end{align}
\end{subequations}
The expressions for higher-order corrections and more complex functions can become increasingly verbose and arcane. Nevertheless, these corrections can be readily computed for any quantity using the method detailed above.
    \section{Model parameters and initial conditions}
        \subsection{Cosmological parameters}
For the purposes of this investigation we used the CLASS Boltzmann software \cite{lesgourgues_cosmic_2011-3} in order to compute the required background and perturbed $\Lambda$CDM quantities. We made use of the default CLASS cosmological paramater values found in the \texttt{explanatory.ini} and \texttt{default.ini} configuration files.
\begin{table}[htb]
\centering
\begin{tabular}{@{}ccccccc@{}}
\toprule\toprule
$h$     & $\Omega_{m0} h^2$ & $T_\text{CMB0}$ & $\Omega_{K0}$ & $10^9 A_s$ & $n_s$     & $k_{\text{piv}}$ \\ \midrule\midrule
0.67810 & 0.1424903      & 2.7255 K       & 0          & 2.100549   & 0.9660499 & 0.05 Mpc$^{-1}$   \\ \bottomrule
\end{tabular}
    \caption{Values of relevant cosmological constants used during this investigation.}
    \label{tbl:Cosmological_Parameters}
\end{table}

A list of the cosmological parameters used in this investigation is shown in  \autoref{tbl:Cosmological_Parameters}. All other relevant $\Lambda$CDM can be inferred from those listed above. In particular, the value of the cosmological constant density parameter today is
\begin{equation}
    \Omega_{\Lambda0} = 1 - \Omega_{m0} - \Omega_{\gamma0} - \Omega_{K0} \approx 0.69\,. 
\end{equation}
In CLASS conventions, the dimensionless primordial power spectrum is given by
\begin{equation}
   \P (k) \equiv P(k)\frac{k^3}{2\pi^2} = A_s\left(\frac{k}{k_{\text{piv}}}\right)^{n_s-1}.
\end{equation}

        \subsection{Anisotropic stress model}\label{sec:Stress_Model_Parameters}

If we align our spatial co-ordinate system with the principal axes of expansion, the spatial shear $\bs{\sigma}$ can be represented as a diagonal matrix
\begin{equation}
    \left(\sigma\indices{^i_j}\right) = \text{diag}(\sigma_1, \sigma_2, -(\sigma_1+\sigma_2)).
\end{equation}
The shear density parameter is then 
\begin{equation}
    \Omega_\sigma = \frac{\sigma^2}{6\H^2} = \frac{1}{3\H^2}\left(\sigma_1^2+\sigma_2^2+\sigma_1\sigma_2\right).
\end{equation}
For a fixed value of $\Omega_\sigma$, the allowed values of the shear parameters lie on an ellipse in the $\sigma_1$-$\sigma_2$ plane. Imposing rotational symmetry about the $z$-axis requires $\sigma_1=\sigma_2$ and hence
\begin{equation}
     \left|\frac{\sigma_1}{\H}\right| = \sqrt{\Omega_\sigma}.
\end{equation}

Recall that the general solution to the shear equation of motion \autoref{eqn:Shear_EoM} for specified value of the shear at time $t_I$ is 
\begin{equation}\label{eqn:Shear_Gen_Sol_Appendix}
    \sigma\indices{^i_j}(a) = \left(\frac{a}{a_I}\right)^{-2}\left[{\sigma_I}\indices{^i_j}+8\pi G a_I^2\int_{a_I}^a\frac{\dd\bar{a}}{a_I} \left(\frac{\bar{a}}{a_I}\right)^{3} \frac{f(\bar{a})}{\H(\bar{a})}W\indices{^i_j}\right].   
\end{equation}
We choose the positive solution to this equation so that $\sigma_1 >0$ and $\sigma_3 < 0$. If the current value of the shear is set to $\bs{\sigma}_0$, the value of the model parameters which satisfies these boundary conditions is
\begin{equation}\label{eqn:W^i_j_Calculation}
    W\indices{^i_j} = \frac{\left({a_0}/{a_I}\right)^{2} {\sigma_0}\indices{^i_j}  - {\sigma_I}\indices{^i_j}}{8\pi G a_I^2\largeint_{\!\!\! a_I}^{a_0}{(\dd\bar{a}}/{a_I}) \left({\bar{a}}/{a_I}\right)^{3} [{f(\bar{a})}/{\H(\bar{a})}]}\,.
\end{equation}
Now, from CMB measurements, we know that the Universe well described by the isotropic FLRW metric throughout much of its history. This means that the ratio $\sigma/\H$ (equivalently $\Omega_\sigma$) must be very small at sufficiently high redshifts. Therefore, if we fix the value of the shear at present and at some high redshift deep in the matter-dominated epoch (say $z\approx 10$), we can determine parameter values $W\indices{^i_j}$ that ensure late-time growth of anisotropy whilst simultaneously eliminating the decaying mode in the late Universe. 

Assuming that the anisotropic stress is negligible during the matter-dominated epoch, we find that the anisotropy ratios at $z_\text{MDE}=10$ and $z_{\text{CMB}} \approx 1100$ (i.e. recombination) are related through
\begin{equation}
    \frac{\sigma_\text{MDE}}{\H_\text{MDE}} = \left(\frac{1+z_\text{MDE}}{1+z_\text{CMB}}\right)^3 \left(\frac{H_\text{CMB}}{H_\text{MDE}}\right)\left(\frac{\sigma_\text{CMB}}{\H_\text{CMB}}\right). 
\end{equation}
Using the constraint $\flatfrac{\sigma_\text{CMB}}{\H_\text{CMB}} \sim 10^{-5}$ and substituting in appropriate values for the Hubble rate we obtain $\flatfrac{\sigma_\text{MDE}}{\H_\text{MDE}} \sim 10^{-8}$ -- i.e. $\Omega_{\sigma,\text{MDE}}\sim 10^{-16}$. We use this value in conjunction with \autoref{eqn:W^i_j_Calculation} to fix model parameters for the anisotropic stress while varying $\bs{\sigma}_0$.

    \section{Details on the computation of angular correlators}\label{sec:Spectra_Details}
        \subsection{\boldmath \texorpdfstring{$E$}{E}- and \texorpdfstring{$B$}{B}-mode auto-correlations}
The first term in \autoref{eqn:E^11E^11_Correlation_Expansion} is simply the ordinary $E$-mode correlation generated by scalar perturbations and scales as $\sim \varphi^2$. Due to statistical isotropy and homogeneity at this level, the correlator is diagonal
\begin{equation}
    \expval{\delta E^{i{\OI}}_{\ell m} \delta E^{j{\OI} *}_{\ell' m'}} = \delta_{\ell \ell'}\delta_{mm'}C^{E^{i}E^{j}}_{\ell}
\end{equation}
and hence is entirely described by the angular power spectrum 
\begin{align}\label{eqn:C^EE_No_Limber}
    C^{E^{i}E^{j}}_{\ell} \equiv \int\dd k\, k^2 P(k) \Delta^{i}_\ell(k)\Delta^{j}_\ell(k),
\end{align}
where we have defined the tomographic kernels
\begin{equation}\label{eqn:01x01_Kernel}
    \Delta^{i}_\ell(k) \equiv \frac{1}{2}\bigg[{\frac{2}{\pi}}\,{\frac{(\ell+2)!}{(\ell-2)!}}\bigg]^{1/2}\int_{0}^{\chi_S}\dd\chi \, q^i(\chi,\chi_S)\,j_{\ell}(k\chi) \, T_\varphi(\chi,k).
\end{equation}

The two ${\II}\times{\OI}$ cross terms in \autoref{eqn:E^11E^11_Correlation_Expansion} scale linearly with the spatial shear as $\sim \varphi^2(\flatfrac{\sigma}{\H})$. A straightforward calculation reveals
\begin{subequations}
    \begin{align}
        \begin{split}    
            \expval{\delta E^{i{\OI}}_{\ell m} \delta E^{j{\II} *}_{\ell' m'}} = \frac{(-1)^{m'}}{2}\left[1+(-1)^{\ell+\ell'}\right] &\sum_{m_1}   \mqty(\ell' & 2 & \ell\\ -m' & m_1 & m) {^{2}F}_{\ell' 2\ell}\\
            & \cdot \int \dd k \,k^2P(k)\Delta^{i}_{\ell}(k)\Delta^{j*}_{\ell m_1}(k)
        \end{split}\label{eqn:E^10E^11_Correlation}\\
        \begin{split}    
            \expval{\delta E^{i{\II}}_{\ell m} \delta E^{j{\OI} *}_{\ell' m'}} = \frac{(-1)^{m}}{2}\left[1+(-1)^{\ell+\ell'}\right] &\sum_{m_1}   \mqty(\ell & 2 & \ell'\\ -m & m_1 & m') {^{2}F}_{\ell 2\ell'}\\
            & \cdot \int \dd k \,k^2P(k)\Delta^{i}_{\ell' m_1}(k)\Delta^{j}_{\ell'}(k),
        \end{split}\label{eqn:E^11E^10_Correlation}
    \end{align}
\end{subequations}

where the isotropic kernel \autoref{eqn:01x01_Kernel} now multiplies the direction-dependent kernel
\begin{equation}
    \Delta^{i}_{\ell m}(k) \equiv \frac{1}{2}\bigg[{\frac{2}{\pi}}\,{\frac{(\ell+2)!}{(\ell-2)!}}\bigg]^{1/2}\int_{0}^{\chi_S}\dd\chi \, q^i(\chi,\chi_S)\,j_{\ell}(k\chi) \,\alpha_{2 m}^{{\IO}}(\chi)\,T_\varphi(\chi,k).
\end{equation}
The selection rule for the 3$j$ symbol (\autoref{eqn:3j_selection_rule}) combined with the alternating factor in \autoref{eqn:E^10E^11_Correlation} and \autoref{eqn:E^11E^10_Correlation} imply that $\ell-\ell' = 0$ or $\ell' -\ell' =  \pm 2$, in agreement with \autoref{eqn:E_E_Selection_Rule}. The final term in the expansion in \autoref{eqn:E^11E^11_Correlation_Expansion} is ${\II}\times{\II}$ and is thus beyond leading order. Putting this together, we find the non-zero BipoSH coefficients
\begin{equation}\label{eqn:E^11E^11_BipoSH}
    \begin{split}
        ^{E^iE^j}\!\mathcal{A}^{LM}_{\ell \ell'} = &(-1)^\ell\sqrt{2\ell+1}\, C^{E^{i}E^{j}}_{\ell}\delta_{\ell \ell'}\delta^{L0}\delta^{M0}\\
        &+ \frac{1}{\sqrt{5}}\left[{^{2}F_{\ell' 2 \ell}}\,\mathcal{C}^{ij}_{\ell M} + (-1)^M\,{^{2}F_{\ell 2 \ell'}}\,\mathcal{C}^{ji*}_{\ell' M}\right]\left(\delta_{\ell \ell'}+ \delta_{\ell\pm 2 ,\ell'}\right)\delta^{L2},
    \end{split}
\end{equation}
where we have defined the object
\begin{equation}
    \mathcal{C}^{ij}_{\ell M} \equiv (-1)^{M}\int \dd k\,k^2P(k)\Delta^{i}_{\ell}(k)\Delta^{j*}_{\ell M}(k)  = \int\dd k \,k^2P(k)\Delta^{i}_{\ell}(k)\Delta^{j}_{\ell, -M}(k).
\end{equation}
We also define the auxiliary quantity
\begin{equation}
    \P^{ij}_{\ell M} \equiv 4 {\frac{(\ell-2)!}{(\ell+2)!}}\, \mathcal{C}^{ij}_{\ell M},
\end{equation}
which strips away the shape factors common to $\Delta_\ell$ and $\Delta_{\ell M}$.

Since it only contains a {\II} contribution, the auto-correlation of the $B$-mode multipoles is 
\begin{equation}
    \begin{split}
        \expval{\delta B^{i{\II}}_{\ell m} \delta B^{j{\II} *}_{\ell' m'}} = \frac{(-1)^{m+m'}}{4} &\sum_{\substack{m_1,m_1' \\ \ell_2,m_2}} \mqty(\ell & 2 & \ell_2\\ -m & m_1 & m_2)\mqty(\ell' & 2 & \ell_2\\ -m' & m_1' & m_2)\\
        &\cdot {^{2}F}_{\ell 2\ell_2}{^{2}F}_{\ell' 2\ell_2}\left[1-(-1)^{\ell+\ell_2}\right]\left[1-(-1)^{\ell'+\ell_2}\right]
        \\
        &\cdot 
        \int {\dd k}\,k^2P(k)\Delta^{i}_{\ell_2 m_1}(k)\Delta^{j*}_{\ell_2 m_1'}(k),
    \end{split}\label{eqn:B^11B^11_Correlation}    
\end{equation}
from which we can calculate the {\biposh} coefficients
\begin{equation}
    \begin{split}
        ^{B^iB^j}\!\A^{LM}_{\ell\ell'} = \frac{1}{4} &\sum_{m,m'}\sum_{\substack{m_1,m_1' \\ \ell_2,m_2}} (-1)^{m'}\mqty(\ell & \ell' & L\\ -m & m' & M) \mqty(\ell & 2 & \ell_2\\ -m & m_1 & m_2)\mqty(\ell' & 2 & \ell_2\\ -m' & m_1' & m_2)\\
        &\cdot {^{2}F}_{\ell 2\ell_2}{^{2}F}_{\ell' 2\ell_2}\left[1-(-1)^{\ell+\ell_2}\right]\left[1-(-1)^{\ell'+\ell_2}\right]\\
        &\cdot \int {\dd k}\,k^2P(k)\Delta^{i}_{\ell_2 m_1}(k)\Delta^{j*}_{\ell_2 m_1'}(k).
    \end{split}\label{eqn:A^BB}    
\end{equation}
The $B$-mode power spectrum is then
\begin{equation}
    C^{B^iB^j}_\ell \equiv (-1)^\ell \frac{^{B^iB^j}\!\A^{00}_{\ell\ell}}{\sqrt{2\ell+1}} = \frac{1}{5}\largesum_{I=\pm 1,m} \frac{\left({^{2}F}_{\ell, 2,\ell+I}\right)^2}{2\ell+1}\int {\dd k}\,k^2P(k)\Delta^{i}_{\ell+I, m}(k)\Delta^{j*}_{\ell+I, m}(k).
\end{equation}

        \subsection{\boldmath \texorpdfstring{$E$}{E}-\texorpdfstring{$B$}{B} cross-correlation}
The leading-order expression for the cross-correlation between the $E$-modes and $B$-modes is almost identical to \autoref{eqn:E^10E^11_Correlation}
\begin{equation}
    \begin{split}\label{eqn:E^10B^11_Correlation}    
        \expval{\delta E^{i{\OI}}_{\ell m} \delta B^{j{\II} *}_{\ell' m'}} = \frac{(-1)^{m'}}{2}\left[1-(-1)^{\ell+\ell'}\right] &\sum_{m_1}   \mqty(\ell' & 2 & \ell\\ -m' & m_1 & m) {^{2}F}_{\ell' 2\ell}\\
        & \cdot \int_{\Reals}k^2P(k)\Delta^{i}_{\ell}(k)\Delta^{j*}_{\ell m_1}(k),
    \end{split}            
\end{equation}
and is only non-zero for $\ell-\ell' = \pm 1$. The corresponding {\biposh} coefficient is therefore
\begin{equation}
    ^{E^iB^j}\!\mathcal{A}^{LM}_{\ell \ell'} = -\frac{i}{4}\frac{(\ell+2)!}{(\ell-2)!}\,\frac{^{2}F_{\ell' 2 \ell}}{\sqrt{5}}\,\P^{ij}_{\ell M}\,\delta^{L2}.
\end{equation}
    \section{Harmonic expansions on the sphere}
Spherical harmonics $Y_{\ell m}$ are special functions defined on the surface of the sphere $S^2$. They are labelled with two integer-valued indices $\ell$ and $m$ which satisfy the relations $\ell \geq 0$ and $-\ell \leq m \leq \ell $, respectively. For a fixed point $(\theta, \phi)$ corresponding to an observation direction $\vb{n}$, they can be expressed as \cite{arfken_mathematical_2005,edmonds_angular_1996}
\begin{equation}\label{eqn;Y_lm_Expression}
    Y_{\ell m}(\vb{n}) = Y_{\ell m}(\theta,\phi) = (-1)^{m}\bigg[{\frac{2\ell+1}{4\pi}\frac{(\ell-m)!}{(\ell+m)!}}\bigg]^{1/2}P_{\ell m}(\cos\theta)e^{im\phi},
\end{equation}
where $P_{\ell m}$ is an associated Legendre polynomial. Spherical harmonics are the eigenfunctions of the (scalar) Laplace operator on the sphere $\laplacian_{S^2}$ and satisfy
\begin{equation}
    \laplacian_{S^2}Y_{\ell m} = -\ell(\ell+1)Y_{\ell m}.
\end{equation}
Spherical harmonics form a complete basis for scalar functions on $S^2$. The completeness and orthogonality properties can be expressed in the form
\begin{subequations}
    \begin{align}
    	\sum_{\ell,m} Y_{\ell m}^*(\vb{n}')Y_{\ell m}(\vb{n})   &= \delta^{(2)}(\vb{n}-\vb{n}')\label{eqn:SH_Completeness}\\
    	\int_{S^2} \dd^2\Omega Y_{\ell m}^*(\vb{n})Y_{\ell' m'}(\vb{n}) &= \delta_{\ell_1\ell_2}\delta_{m_1m_2}\label{eqn:SH_Normalisation},
    \end{align}
\end{subequations}
where $\delta^{(2)}(\vb{n}-\vb{n}')=\delta(\phi-\phi')\delta(\cos\theta-\cos\theta')$ and $\dd^2\Omega = \dd \theta \sin\theta\dd\phi$ are the two-dimensional Dirac delta function and the volume element on the sphere, respectively. The completeness and normalisation of spherical harmonics allows us to expand a scalar function $f$ on the sphere as
\begin{equation}\label{eqn:f_expansion}
	f(\vb{n}) = \sum_{\ell,m}f_{\ell m}Y_{\ell m}(\vb{n}),
\end{equation}
where the multipole coefficients are given by
\begin{equation}\label{eqn:f_multipoles}
	f_{\ell m} = \int_{S^2} \dd^2\Omega \,Y_{\ell m}^*(\vb{n})f(\vb{n}).
\end{equation}
The complex conjugate of a spherical harmonic flips the $m$ index and introduces a phase factor
\begin{equation}\label{eqn:SH_Conjugate}
    Y_{\ell m}(\vb{n})^{*} = Y_{\ell m}(\theta,-\phi) = (-1)^{m}Y_{\ell (-m)}(\vb{n}).
\end{equation}
Under the parity inversion $\vb{n}\longmapsto -\vb{n}$
\begin{align}\label{eqn:SH_Parity}
    Y_{\ell m}(-\vb{n}) = Y_{\ell m}(\pi-\theta,\phi+\pi) = (-1)^{\ell}Y_{\ell m}(\vb{n}),
\end{align}
and hence that for even (odd) $\ell$, $Y_{\ell m}$ is parity even (odd).

        \subsection{Spin-weighted spherical harmonics}
`Spin', `spin weight', or `helicity' is a property of certain functions on $S^2$ related to their behaviour under three-dimensional rotations. Consider a counter-clockwise rotation by an angle $\alpha$ about the axis $\vb{n}$. Under this mapping, an object which transforms according to 
\begin{equation}
    f^{s}(\vb{n}) \longmapsto e^{is\alpha} f^{s}(\vb{n}),
\end{equation}
where $s \in \Ints$, is said to be of spin $|s|$ or helicity $s$ \cite{durrer_cosmic_2020}. Scalar functions are necessarily spin-0 due to their simple transformation properties. Moreover, ordinary spherical harmonics are used to expand scalar functions and must therefore inherit their spin. Objects with non-zero spin are actually (components of) tensor-valued functions on $S^2$ and are only defined relative to a certain co-ordinate basis. Any symmetric and traceless rank-2 tensor field (like polarisation or lensing shear) is automatically spin-2. Geometrically, this is because -- when viewed as linear operators -- they have the effect of distorting circles into ellipses and thus have $180^\circ$ rotational symmetry. Objects of a particular spin need to be expanded in harmonics which respect their rotational symmetries.

The spin raising-operator $\eth$ and the spin-lowering operator $\ethb$ have the effect of increasing and decreasing an object's spin, respectively. Written in terms of the spherical co-ordinates $\theta$ and $\phi$, the action of these operators on a spin-$s$ function is
\begin{subequations}
    \begin{align}
        \tilde{f}^{s+1} &\equiv \eth f^{s}  = -\left(\partial_\theta +i\csc\theta\partial_\phi -s\cot\theta\right)f^{s}\\
        \tilde{f}^{s-1} &\equiv \ethb f^{s} = -\left(\partial_\theta -i\csc\theta\partial_\phi +s\cot\theta\right)f^{s}.
    \end{align}
\end{subequations}
We can therefore construct harmonics of higher spin through successive actions of these operators on ordinary (spin-0) spherical harmonics. In analogy with ladder operators of quantum mechanics, the spin-weighted spherical harmonic (SWSH) $Y_{\ell m}^{s}$ of spin-weight $s$ is given by \cite{newman_note_1966,goldberg_spin_1967,boyle_how_2016}
\begin{equation} \label{eqn:SWSH from SH}
    Y^s_{\ell m}  =\begin{cases}  \hphantom{(-1)^s}\sqrt{{(\ell+s)!}/{(\ell-s)!}}~\eth^sY^\ell_m  &\text{ for } ~~~0<s\leq \ell \\
     (-1)^s\sqrt{{(\ell-s)!}/{(\ell+s)!}}~\ethb^{|s|}Y^\ell_m  &\text{ for } -\ell\leq s<0.
                   \end{cases}  	
\end{equation}
Like normal spherical harmonics, the SWSHs are eigenfunctions of the Laplace operator on the sphere, albeit with an altered eigenvalue \cite{hu_weak_2000}
\begin{equation}\label{eqn:SWSH_E-val}
	\laplacian_{S^2}Y^{s}_{\ell m} = -\left[\ell(\ell+1)-s^2\right]Y^{s}_{\ell m}.
\end{equation}
For fixed $s$, SWSHs are orthogonal
\begin{equation}
	\int_{S^2} \dd^2\Omega\, Y^{s*}_{\ell m}(\vb{n})Y^{s}_{\ell' m'}(\vb{n}) = \delta_{\ell \ell'}\delta_{m m'}
\end{equation}
and form a complete basis for objects of their particular spin. A spin-$s$ quantity $f^{s}$ can therefore be expanded in terms of spin-$s$ harmonics as
\begin{equation}
	f^{s}(\vb{n}) = \sum_{\ell,m} {f^{s}_{\ell m}} Y^s_{\ell m}(\vb{n}),
\end{equation}
where the multipoles $f^{s}_{\ell m}$ are given by
\begin{equation}
	f^{s}_{\ell m} = \int_{S^{2}}\dd^2\Omega\, Y^{s*}_{\ell m}(\vb{n})f^{s}(\vb{n}).
\end{equation}
Under complex conjugation, \autoref{eqn:SH_Conjugate} becomes
\begin{equation}\label{eqn:SWSH_Conjugate}
    Y^{s*}_{\ell m}(\vb{n}) = (-1)^{s+m}Y^{-s}_{\ell (-m)}(\vb{n})
\end{equation}
while the parity relation \autoref{eqn:SH_Parity} generalises to
\begin{equation}\label{eqn:SWSH_Parity}
    Y^s_{\ell m}(-\vb{n}) = (-1)^{\ell}Y^{-s}_{\ell m}(\vb{n}).
\end{equation}

For the particular case of the spin-2 shear variable $\gamma_1 \pm i\gamma_2$, the multipole decomposition reads
\begin{equation}
    \gamma_1(\vb{n})  \pm i\gamma_2(\vb{n}) = \sum_{\ell,m}(E_{\ell m} \pm i B_{\ell m}) Y^{\pm 2}_{\ell m}(\vb{n}).
\end{equation}
Using \autoref{eqn:SWSH_Conjugate}, we see that $E^*_{\ell m}=(-1)^mE_{\ell(-m)}$ and $B^*_{\ell m}=(-1)^mB_{\ell(-m)}$, as with real scalar variables. Furthermore, \autoref{eqn:SWSH_Parity} and the parity relation $[\gamma_1  \pm i\gamma_2](\vb{n})\longmapsto [\gamma_1  \mp i\gamma_2](-\vb{n})$ imposes the transformation properties $E_{\ell m}\longmapsto (-1)^\ell E_{\ell m}$ and $B_{\ell m}\longmapsto (-1)^{\ell+1} B_{\ell m}$ on the $E$-modes and $B$-modes, respectively.

        \subsection{\boldmath Wigner 3\texorpdfstring{$j$}{j} symbols}
Integrals of the product of three SWSHs arise frequently in the study of anisotropy. The {Gaunt integral} \cite{edmonds_angular_1996}
\begin{equation}
	\begin{split}
	\int_{S^2}\dd^2\Omega\, Y^{s_1}_{\ell_1 m_1}(\vb{n})Y^{s_2}_{\ell_2 m_2}(\vb{n})Y^{s_3}_{\ell_3 m_3}(\vb{n}) = &\bigg[{\frac{(2\ell_1+1)(2\ell_2+1)(2\ell_2+1)}{4\pi}}\bigg]^{1/2}\\ &\cdot\mqty(\ell_1 & \ell_2 & \ell_3\\ -s_1 & -s_2 & -s_3)\mqty(\ell_1 & \ell_2 & \ell_3\\ m_1 & m_2 & m_3)
	\end{split}
\end{equation}
expresses integrals of this type in terms of {Wigner 3j symbols}. These objects are only non-zero if its elements satisfy the selection rules
\begin{equation}\label{eqn:3j_selection_rule}
	m_1+m_2+m_3=0 \quad \text{and} \quad |\ell_i-\ell_j| \leq \ell_{k} \leq \ell_{i}+\ell_{j}.
\end{equation}
They also satisfy the following symmetries
\begin{subequations}
\begin{enumerate}
	\item Even permutation of columns:
\end{enumerate}
	\begin{equation}
		\mqty(\ell_1 & \ell_2 & \ell_3\\
			  m_1    & m_2    & m_3) = 
		\mqty(\ell_3 & \ell_1 & \ell_2\\
			  m_3    & m_1    & m_2) =
		\mqty(\ell_2 & \ell_3 & \ell_1\\
			  m_2    & m_3    & m_1)	  
	\end{equation}
\begin{enumerate}[resume]
	\item Odd permutation of columns:
\end{enumerate}
	\begin{equation}
		\mqty(\ell_1 & \ell_2 & \ell_3\\
			  m_1    & m_2    & m_3) = 
		(-1)^{\ell_1+\ell_2+\ell_3}\mqty(\ell_2 & \ell_1 & \ell_3\\
			  m_2    & m_1    & m_3) =
		(-1)^{\ell_1+\ell_2+\ell_3}\mqty(\ell_1 & \ell_3 & \ell_2\\
			  m_1    & m_3    & m_1)	  
	\end{equation}
\begin{enumerate}[resume]
	\item Negation of second row:
\end{enumerate}
\begin{equation}\label{eqn:3j_neg_prop}
	\mqty(\ell_1 & \ell_2 & \ell_3\\
			  m_1    & m_2    & m_3) = 
		(-1)^{\ell_1+\ell_2+\ell_3}\mqty(\ell_1 & \ell_2 & \ell_3\\
			  -m_1    & -m_2    & -m_3).
\end{equation}
\end{subequations}
Moreover, the 3j symbols possess the useful orthogonality properties
\begin{subequations}
	\begin{align}
		\sum_{m_1,m_2}\mqty(\ell_1 & \ell_2 & \ell\\
			  m_1    & m_2    & m)
			  \mqty(\ell_1 & \ell_2 & \ell'\\
			  m_1    & m_2    & m') &= \frac{1}{2\ell+1}\delta_{\ell \ell'}\delta_{m m'}\label{eqn:3j_Orthog_1}\\
		\sum_{\ell,m}(2\ell+1)\mqty(\ell_1 & \ell_2 & \ell\\
			  m_1    & m_2    & m)
			  \mqty(\ell_1 & \ell_2 & \ell\\
			  m_1'    & m_2'    & m) &= \delta_{m_1 m_1'}\delta_{m_2 m_2'}.\label{eqn:3j_Orthog_2}
	\end{align}		
\end{subequations}

Following \cite{hu_weak_2000}, we define the integral
\begin{equation}
	^sI^{m_1m_2m_3}_{\ell_1\ell_2\ell_3} \equiv \int_{S^2}\dd^2\Omega \,Y^{s*}_{\ell_1m_1}(\vb{n})\left[D^AY_{\ell_2m_2}(\vb{n})\right]\left[D_AY^{s}_{\ell_3m_3}(\vb{n})\right],
\end{equation}
which appears in post-Born corrections to lensing variables and lensed CMB power spectra. By making use of the product rule for derivatives and defining the symbol \cite{hu_weak_2000,pitrou_weak-lensing_2015}
\begin{equation}\label{eqn:sF_ell_1ell_2ell_3_defn}
	\begin{split}
	^sF_{\ell_1\ell_2\ell_3} \equiv & \frac{1}{2}\big[\ell_2(\ell_2+1)+\ell_3(\ell_3+1)-\ell_1(\ell_1+1)\big]\\
	&\cdot\bigg[{\frac{(2\ell_1+1)(2\ell_2+1)(2\ell_3+1)}{4\pi}}\bigg]^{1/2}\mqty(\ell_1 & \ell_2 & \ell_3\\ s & 0 & -s),
	\end{split}
\end{equation}
we can separate the parts of $^sI^{m_1m_2m_3}_{\ell_1\ell_2\ell_3}$ that depend on $s$ from those that depend on $m_1,m_2$, and $m_3$
\begin{equation}
	^sI^{m_1m_2m_3}_{\ell_1\ell_2\ell_3} = {^sF}_{\ell_1\ell_2\ell_3}(-1)^{m_1+s}\mqty(\ell_1 & \ell_2 & \ell_3\\ -m_1 & m_2 & m_3).
\end{equation}
Using \autoref{eqn:3j_neg_prop}, we see that both of these quantities inherit the property
\begin{subequations}
	\begin{align}
		 {^{-s}I}^{m_1m_2m_3}_{\ell_1\ell_2\ell_3}&= (-1)^{\ell_1+\ell_2+\ell_3}   {^sI}^{m_1m_2m_3}_{\ell_1\ell_2\ell_3}\\
		 {^{-s}F}_{\ell_1\ell_2\ell_3}&= (-1)^{\ell_1+\ell_2+\ell_3}  {^sF}_{\ell_1\ell_2\ell_3}.\label{eqn:sF_ell_1ell_2ell_3_neg}
	\end{align}
\end{subequations}
Four specific sets of values of $^sF_{\ell_1\ell_2\ell_3}$ -- which emerge in {\II} corrections to the $B$-mode shear multipoles -- are given by Mathematica as
\begin{subequations}
	\begin{align}
        ^2F_{\ell+1, 2, \ell} &= (-1)^{\ell+1}(\ell-2)\bigg[{\frac{15}{\pi}}\,{\frac{(\ell-1)(\ell+3)}{\ell(\ell+1)(\ell+2)}}\bigg]^{1/2} \\
        ^2F_{\ell-1, 2, \ell} &= (-1)^{\ell+1}(\ell+3)\bigg[{\frac{15}{\pi}}\,{\frac{(\ell-2)(\ell+2)}{\ell(\ell-1)(\ell+1)}}\bigg]^{1/2}\\
		^2F_{\ell, 2, \ell+1} &= (-1)^{\ell}(\ell+4)\bigg[{\frac{15}{\pi}}\,{\frac{(\ell-1)(\ell+3)}{\ell(\ell+1)(\ell+2)}}\bigg]^{1/2} \\
		^2F_{\ell, 2, \ell-1} &= (-1)^{\ell}(\ell-3)\bigg[{\frac{15}{\pi}}\,{\frac{(\ell-2)(\ell+2)}{\ell(\ell-1)(\ell+1)}}\bigg]^{1/2}.
	\end{align}
\end{subequations}

        \subsection{Bipolar spherical harmonics}\label{sec:BipoSH_Properties}
Departures from statistical isotropy induce correlations between different scales which cannot be captured by standard angular power spectra. Bipolar spherical harmonics (\biposh s) are a particular class of special function that can be used to study the tell-tale signatures of various sources of anisotropy.

We define the bipolar spherical harmonic as a particular linear combination of Wigner 3$j$ symbols and products of spherical harmonics \cite{hajian_cosmic_2005}
\begin{equation}
    \left\{Y_\ell(\vb{n})\otimes Y_{\ell'}(\vb{n}')\right\}_{LM} \equiv  (-1)^{\ell-\ell'+M}\sqrt{2L+1}\sum_{m,m'} \mqty(\ell & \ell' & L \\ m & m' & -M) Y_{\ell m}(\vb{n})Y_{\ell' m'}(\vb{n}'). 
\end{equation}
{\biposh}s form a complete set of basis functions on $S^2\times S^2$. Moreover, they are orthonormal
\begin{equation}
    %\begin{split}    
    \int_{S^2}\!\dd\Omega\!\int_{S^2}\!\dd\Omega' \left\{Y_{\ell_1}(\vb{n})\otimes Y_{\ell_1'}(\vb{n}')\right\}^{*}_{L_1M_1}\! \left\{Y_{\ell_2}(\vb{n})\otimes Y_{\ell_2'}(\vb{n}')\right\}_{L_2M_2} = \delta_{\ell_1\ell_2}\delta_{\ell_1'\ell_2'} \delta_{L_1L_2}\delta_{M_1M_2}\,,
   % \end{split}
\end{equation}
and its indices \{$\ell$, $\ell'$, $L$\} must satisfy the triangle inequality $|\ell-\ell'|\leq L \leq \ell +\ell'$.

Without any assumptions on statistical symmetry, the real-space correlation of the stochastic variables $X$ and $Z$
\begin{equation}
    C^{XZ}(\vb{n}, \vb{n}') = \expval{X(\vb{n})Z^*(\vb{n}')}
\end{equation}
is a two-point function defined on $S^2\times S^2$. This bivariate function can be written as the linear combination
\begin{align}
    C^{XZ}(\vb{n}, \vb{n}') = \sum_{\substack{\ell,m,\ell',m' \\ L,M}} {^{XZ}\!\A^{LM}_{\ell\ell'}} \left\{Y_\ell(\vb{n})\otimes Y_{\ell'}(\vb{n}')\right\}_{LM} 
\end{align}
where the {\biposh} coefficients ${^{XZ}\!\A^{LM}_{\ell\ell'}}$ are given by
\begin{equation}
    {^{XZ}\!\A^{LM}_{\ell\ell'}} = \int_{S^2}\dd\Omega\int_{S^2}\dd\Omega'\left\{Y_\ell(\vb{n})\otimes Y_{\ell'}(\vb{n}')\right\}_{LM}^{*} C^{XZ}(\vb{n}, \vb{n}').
\end{equation}
Written in terms of the correlations between the multipoles $X_{\ell m}$ and $Z_{\ell' m'}$, this becomes \cite{hajian_cosmic_2005}
\begin{equation}\label{eqn:BipoSH_Coeff_Defn_XZ}
    ^{XZ}\!\A^{LM}_{\ell \ell'} = \sqrt{2L+1}\largesum_{m,m'}(-1)^{L+m} \mqty(\ell & \ell' & L \\ -m & m' & M) \expval{X_{\ell m}Z^*_{\ell' m'}}. 
\end{equation}

Suppose the multipoles $\hat{X}_{\ell m}$ and $\hat{Z}_{\ell' m'}$ have been determined from some sky map. An un-biased statistical estimator for the corresponding {\biposh} coefficient for this map is then
\begin{equation}\label{eqn:BipoSH_Coeff_Estimator_XZ}
    {^{XZ}\!\hat{\A}}^{LM}_{\ell \ell'} = \sqrt{2L+1}\largesum_{m,m'}(-1)^{L+m} \mqty(\ell & \ell' & L \\ -m & m' & M) \hat{X}_{\ell m}\hat{Z}^*_{\ell' m'}. 
\end{equation}
Under the assumption of statistical isotropy, the auto- and cross-correlations between the multipoles are diagonal
\begin{subequations}
    \begin{align}
        \expval{\hat{X}_{\ell m} \hat{X}^{*}_{\ell' m'}}_{\text{SI}} &\equiv {\left(C^{XX}_{\ell}\right)_{\text{SI}}} \delta_{\ell \ell'}\delta_{m m'} \\
        \expval{\hat{Z}_{\ell m} \hat{Z}^{*}_{\ell' m'}}_{\text{SI}} &\equiv {\left(C^{ZZ}_{\ell}\right)_{\text{SI}}}\delta_{\ell \ell'}\delta_{m m'}  \\
        \expval{\hat{X}_{\ell m} \hat{Z}^{*}_{\ell' m'}}_{\text{SI}} &\equiv {\left(C^{XZ}_{\ell}\right)_{\text{SI}}}\delta_{\ell \ell'}\delta_{m m'}.
    \end{align}
\end{subequations}
The expected value of the estimator in \autoref{eqn:BipoSH_Coeff_Estimator_XZ} is thus
\begin{equation}
    \expval{{^{XZ}\!\hat{\A}}^{LM}_{\ell \ell'}}_{\text{SI}} = (-1)^{\ell}\sqrt{2\ell+1} {\left(C^{XZ}_{\ell}\right)_{\text{SI}}} \delta_{\ell \ell'}\delta_{L0}\delta_{M0}.
\end{equation}
The (statistically isotropic) covariance between two different {\biposh} coefficients is defined as
\begin{equation}\label{eqn:BipoSH_SI_Cov_Defn}
    \begin{split}    
        \text{Cov}\left({^{X_1 Z_1}\!\hat{\A}^{L_1M_1}_{\ell_1 \ell_1'}},{^{X_2 Z_2}\!\hat{\A}^{L_2M_2}_{\ell_2 \ell_2'}}\right)_\text{SI} \equiv &\expval{{^{X_1 Z_1}\!\hat{\A}^{L_1M_1}_{\ell_1 \ell_1'}} \cdot {^{X_2 Z_2}\!\hat{\A}^{L_2M_2*}_{\ell_2 \ell_2'}} }_\text{SI}\\
        &- \expval{{^{X_1 Z_1}\!\hat{\A}^{L_1M_1}_{\ell_1 \ell_1'}} }_\text{SI}\expval{{^{X_2 Z_2}\!\hat{\A}^{L_2M_2 *}_{\ell_2 \ell_2'}} }_\text{SI}       
    \end{split}
\end{equation}
If we further assume that the stochastic variables $X_1$, $X_2$, $Z_1$, and $Z_2$ are Gaussian, applying Wick's/Isserlis's theorem reduces this expression to
\begin{equation}\label{eqn:BipoSH_SI_Cov_General}
    \begin{split}
        \text{Cov}\Bigl(  \cdots \Bigr)_\text{SI} = &\left[  {\left(C^{X_1X_2}_{\ell_1}\right)_{\text{SI}}} {\left(C^{Z_1Z_2}_{\ell_1'}\right)_{\text{SI}}} \delta_{\ell_1\ell_2} \delta_{\ell_1'\ell_2'}\right.\\
        &+ \left.  (-1)^{L + \ell_1 + \ell_1'}  {\left(C^{X_1Z_2}_{\ell_1}\right)_{\text{SI}}} {\left(C^{Z_1X_2}_{\ell_1'}\right)_{\text{SI}}} \delta_{\ell_1\ell_2'} \delta_{\ell_1'\ell_2}      \right]\delta^{L_1L_2} \delta^{M_1M_2}.
    \end{split}
\end{equation}
In \S \ref{sec:Estimator_SNR} we make use of this expression in order to calculate approximate signal-to-noise ratios for the estimator $\hat{\P}^{ij}_{\ell M}$.
%%%%%%%%%%%%%%%%%%%%%%%%%%%%%%%%%%%%%%%%%%%%%%%%%%%%%%%%%%%%%%%%%%%%%%%%%%%%%%%%%%%%%%%%%%%%%%%%%%%%%%%%
    \section{Computation of tomographic source distributions}
    \label{sec:sourcedistrib}
For Euclid, the underlying source distribution is modelled as \cite{blanchard_euclid_2020, deshpande_euclid_2024} 
\begin{equation}\label{eqn:n_underlying}
	n(z)\propto \left(\frac{z}{z_0}\right)^2 \exp\left[-\left(\frac{z}{z_0}\right)^\frac{3}{2}\right],
\end{equation}
where $z_0=z_\text{m}/\sqrt{2}$ and the median redshift $z_\text{m}=0.9$. Euclid uses 10 equi-populated bins with edges $z_i^{\pm} \in$ \{0.001, 0.42, 0.56, 0.68, 0.79, 0.90, 1.02, 1.15, 1.32, 1.58, 2.50\}. The tomographic galaxy source distribution for the $i$\text{th} redshift bin is given by 
\begin{equation}\label{eqn:n^i_Definition}
n_i(z)=\frac{\largeint_{z_i^-}^{z_i^+}\dd z_\text{p}n(z) p_\text{ph}(z_\text{p}|z)}{\largeint_{z_\text{min}}^{z_\text{max}}\dd z\largeint_{z_i^-}^{z_i^+}\text{d}z_\text{p}n(z)p_\text{ph}(z_\text{p}|z)},
\end{equation}
where ${z_i^-}$ and ${z_i^+}$ define the edges of the bin, and $(z_\text{min},z_\text{max}) = ({z_1^-}, {z_{10}^+}) = (0.001, 2.5)$. The full source distribution $n_\text{F}(z)$ is obtained by replacing the integral limits $({z_i^-}, {z_{i}^+})\mapsto (z_\text{min},z_\text{max})$ so that the entire redshift range is taken into account. In \autoref{eqn:n^i_Definition}, the underlying distribution $n(z)$ has been convolved with the probability distribution function $p_\text{ph}(z_\text{p}|z)$ describing the probability that a galaxy with redshift $z$ has a measured redshift $z_\text{p}$. The probability distribution is modelled as
\begin{equation}
    \begin{split}
        p_{\text{ph}}(z_\text{p}|z)&= \frac{1-f_{\text{out}}}{\sqrt{2\pi}\sigma_{\text{b}}(1+z)}\exp\left\{-\frac{1}{2}\left[\frac{z-c_\text{b}z_\text{p}-z_\text{b}}{\sigma_{\text{b}}(1+z)}\right]^{2}\right\} \\ &+\frac{f_{\text{out}}}{\sqrt{2\pi}\sigma_{\text{o}}(1+z)}\exp\left\{-\frac{1}{2}\left[\frac{z-c_\text{o}z_\text{p}-z_\text{o}}{\sigma_{\text{o}}(1+z)}\right]^{2}\right\},
    \end{split}
\end{equation}
with $(c_\text{b}, z_\text{b}, \sigma_\text{b}, c_\text{o}, z_\text{o}, \sigma_\text{o}, f_\text{out})=(1.0,0.0,0.05,1.0, 0.1, 0.05, 0.1)$. The result of this convolution process is shown in  \autoref{fig:Euclid_Tomographic_Distributions}.

%%%%%%%%%%%%%%%%%%%%%%%%%%%%%%%%%
\newpage
\cleardoublepage

\bibliographystyle{JHEP}
\bibliography{B-Modes_References}

\end{document}

%% file: Packages_Macros.tex
\usepackage{jcappub}
\usepackage{multicol}
\usepackage{indentfirst}

\usepackage[UKenglish]{babel}
\usepackage[utf8]{inputenc}
\usepackage[T1]{fontenc}
\usepackage{amsfonts}
\usepackage{verbatim}
\usepackage{hyphenat}

\usepackage{graphicx}
\usepackage{float}
\usepackage{epsfig}
\usepackage{rotate}
\usepackage{caption}
\usepackage{booktabs}
\usepackage{tablefootnote}
\usepackage{afterpage}

\usepackage{amsmath}
\usepackage{amssymb}
\usepackage{mathrsfs}
\usepackage{empheq}
\usepackage{relsize}
\usepackage{bm}
\allowdisplaybreaks

\usepackage{physics}
\usepackage{tensor}

\usepackage{xcolor}
\usepackage[colorinlistoftodos, tickmarkheight=0.2cm, textsize=tiny]{todonotes}
  
\usepackage{enumitem}
\numberwithin{equation}{section}
\numberwithin{figure}{section}
\numberwithin{table}{section}

\newcommand{\bs}[1]{\boldsymbol{#1}}

\renewcommand{\H}{\mathcal{H}}

\newcommand{\N}{\mathcal{N}}
\renewcommand{\P}{\mathcal{P}}

\DeclareMathAlphabet\mathbfcal{OMS}{cmsy}{b}{n}
\newcommand{\A}{\mathbfcal{A}}

\newcommand{\D}{\mathbfcal{D}}

\newcommand{\Reals}[1][]{\mathbb{R}^{#1}}
\newcommand{\Ints}{\mathbb{Z}}

\newcommand{\OO}{\{0,0\}}
\newcommand{\IO}{\{1,0\}}
\newcommand{\OI}{\{0,1\}}
\newcommand{\II}{\{1,1\}}

\newcommand{\biposh}{BipoSH}

\newcommand{\largesum}{\mathop{\mathlarger{\sum}}}
\newcommand{\largeint}{\mathop{\mathlarger{\int}}}
\newcommand{\ethb}{\bar{\eth}}

%% file: main.bbl
\providecommand{\href}[2]{#2}\begingroup\raggedright\begin{thebibliography}{100}

\bibitem{Aluri:2022hzs}
P.K.~Aluri et~al., \emph{{Is the observable Universe consistent with the cosmological principle?}}, \href{https://doi.org/10.1088/1361-6382/acbefc}{\emph{Class. Quant. Grav.} {\bfseries 40} (2023) 094001} [\href{https://arxiv.org/abs/2207.05765}{{\ttfamily 2207.05765}}].

\bibitem{Euclid:2022ucc}
{\scshape Euclid} collaboration, \emph{{Euclid: Testing the Copernican principle with next-generation surveys}}, \href{https://doi.org/10.1051/0004-6361/202244557}{\emph{Astron. Astrophys.} {\bfseries 671} (2023) A68} [\href{https://arxiv.org/abs/2207.09995}{{\ttfamily 2207.09995}}].

\bibitem{bunn_how_1996}
E.F.~Bunn, P.~Ferreira and J.~Silk, \emph{How {{Anisotropic}} is our {{Universe}}?}, \href{https://doi.org/10.1103/PhysRevLett.77.2883}{\emph{Phys. Rev. Lett.} {\bfseries 77} (1996) 2883} [\href{https://arxiv.org/abs/astro-ph/9605123}{{\ttfamily astro-ph/9605123}}].

\bibitem{kogut_limits_1997}
A.~Kogut, G.~Hinshaw and A.J.~Banday, \emph{Limits to {{Global Rotation}} and {{Shear From}} the {{COBE DMR}} 4-{{Year Sky Maps}}}, \href{https://doi.org/10.1103/PhysRevD.55.1901}{\emph{Phys. Rev. D} {\bfseries 55} (1997) 1901} [\href{https://arxiv.org/abs/astro-ph/9701090}{{\ttfamily astro-ph/9701090}}].

\bibitem{martinez-gonzalez_delta_1995}
E.~Martinez-Gonzalez and J.L.~Sanz, \emph{{$\Delta T/T$}and the isotropy of the universe.}, {\emph{Astronomy and Astrophysics} {\bfseries 300} (1995) 346}.

\bibitem{maartens_anisotropy_1996}
R.~Maartens, G.~Ellis and W.~Stoeger, \emph{Anisotropy and inhomogeneity of the universe from {{$\Delta$T}}/{{T}}}, \href{https://doi.org/10.48550/arXiv.astro-ph/9510126}{\emph{Astron. Astrophys.} {\bfseries 309} (1996) L7} [\href{https://arxiv.org/abs/astro-ph/9510126}{{\ttfamily astro-ph/9510126}}].

\bibitem{saadeh_how_2016}
D.~Saadeh, S.M.~Feeney, A.~Pontzen, H.V.~Peiris and J.D.~McEwen, \emph{How isotropic is the {{Universe}}?}, \href{https://doi.org/10.1103/PhysRevLett.117.131302}{\emph{Phys. Rev. Lett.} {\bfseries 117} (2016) 131302} [\href{https://arxiv.org/abs/1605.07178}{{\ttfamily 1605.07178}}].

\bibitem{akarsu_testing_2023}
O.~Akarsu, E.~Di~Valentino, S.~Kumar, M.~Ozyigit and S.~Sharma, \emph{Testing spatial curvature and anisotropic expansion on top of the {$\Lambda$CDM} model}, \href{https://doi.org/10.1016/j.dark.2022.101162}{\emph{Phys. Dark Univ.} {\bfseries 39} (2023) 101162} [\href{https://arxiv.org/abs/2112.07807}{{\ttfamily 2112.07807}}].

\bibitem{anninos_how_1991}
P.~Anninos, R.A.~Matzner, T.~Rothman and M.P.~Ryan, \emph{How does inflation isotropize the {{Universe}}?}, \href{https://doi.org/10.1103/PhysRevD.43.3821}{\emph{Phys. Rev. D} {\bfseries 43} (1991) 3821}.

\bibitem{pitrou_predictions_2008}
C.~Pitrou, T.S.~Pereira and J.-P.~Uzan, \emph{Predictions from an anisotropic inflationary era}, \href{https://doi.org/10.1088/1475-7516/2008/04/004}{\emph{JCAP} {\bfseries 04} (2008) 004} [\href{https://arxiv.org/abs/0801.3596}{{\ttfamily 0801.3596}}].

\bibitem{rothman_effects_1982}
T.~Rothman and R.~Matzner, \emph{Effects of {{Anisotropy}} and {{Dissipation}} on the {{Primordial Light-Isotope Abundances}}}, \href{https://doi.org/10.1103/PhysRevLett.48.1565}{\emph{Phys. Rev. Lett.} {\bfseries 48} (1982) 1565}.

\bibitem{rothman_nucleosynthesis_1984}
T.~Rothman and R.~Matzner, \emph{Nucleosynthesis in anisotropic cosmologies revisited}, \href{https://doi.org/10.1103/PhysRevD.30.1649}{\emph{Phys. Rev. D} {\bfseries 30} (1984) 1649}.

\bibitem{matzner_conjecture_1986}
R.~Matzner, T.~Rothman and G.F.R.~Ellis, \emph{Conjecture on isotope production in the {{Bianchi}} cosmologies}, \href{https://doi.org/10.1103/PhysRevD.34.2926}{\emph{Phys. Rev. D} {\bfseries 34} (1986) 2926}.

\bibitem{campanelli_helium-4_2011}
L.~Campanelli, \emph{Helium-4 {{Synthesis}} in an {{Anisotropic Universe}}}, \href{https://doi.org/10.1103/PhysRevD.84.123521}{\emph{Phys. Rev. D} {\bfseries 84} (2011) 123521} [\href{https://arxiv.org/abs/1112.2076}{{\ttfamily 1112.2076}}].

\bibitem{barrow_limits_1997}
J.D.~Barrow, \emph{Limits on cosmological magnetic fields and other anisotropic stresses},  in \emph{4th {{Paris Cosmology Colloquium}}}, pp.~196--222, June, 1997 [\href{https://arxiv.org/abs/gr-qc/9712020}{{\ttfamily gr-qc/9712020}}].

\bibitem{sharif_dynamics_2010}
M.~Sharif and M.~Zubair, \emph{Dynamics of {{Bianchi I Universe}} with {{Magnetized Anisotropic Dark Energy}}}, \href{https://doi.org/10.1007/s10509-010-0414-y}{\emph{Astrophys. Space Sci.} {\bfseries 330} (2010) 399} [\href{https://arxiv.org/abs/1005.4480}{{\ttfamily 1005.4480}}].

\bibitem{koivisto_accelerating_2008}
T.~Koivisto and D.F.~Mota, \emph{Accelerating {{Cosmologies}} with an {{Anisotropic Equation}} of {{State}}}, \href{https://doi.org/10.1086/587451}{\emph{Astrophys. J.} {\bfseries 679} (2008) 1} [\href{https://arxiv.org/abs/0707.0279}{{\ttfamily 0707.0279}}].

\bibitem{koivisto_anisotropic_2008}
T.~Koivisto and D.F.~Mota, \emph{Anisotropic {{Dark Energy}}: {{Dynamics}} of {{Background}} and {{Perturbations}}}, \href{https://doi.org/10.1088/1475-7516/2008/06/018}{\emph{JCAP} {\bfseries 06} (2008) 018} [\href{https://arxiv.org/abs/0801.3676}{{\ttfamily 0801.3676}}].

\bibitem{akarsu_lrs_2010}
O.~Akarsu and C.B.~Kilinc, \emph{{{LRS Bianchi Type I Models}} with {{Anisotropic Dark Energy}} and {{Constant Deceleration Parameter}}}, \href{https://doi.org/10.1007/s10714-009-0821-y}{\emph{Gen. Rel. Grav.} {\bfseries 42} (2010) 119} [\href{https://arxiv.org/abs/0807.4867}{{\ttfamily 0807.4867}}].

\bibitem{akarsu_bianchi_2010}
O.~Akarsu and C.B.~Kilinc, \emph{Bianchi type {{III}} models with anisotropic dark energy}, \href{https://doi.org/10.1007/s10714-009-0878-7}{\emph{Gen. Rel. Grav.} {\bfseries 42} (2010) 763} [\href{https://arxiv.org/abs/0909.1025}{{\ttfamily 0909.1025}}].

\bibitem{akarsu_sitter_2010}
O.~Akarsu and C.B.~Kilinc, \emph{De {{Sitter}} expansion with anisotropic fluid in {{Bianchi}} type-{{I}} space-time}, \href{https://doi.org/10.1007/s10509-009-0254-9}{\emph{Astrophys. Space Sci.} {\bfseries 326} (2010) 315} [\href{https://arxiv.org/abs/1001.0550}{{\ttfamily 1001.0550}}].

\bibitem{kumar_bianchi_2012}
S.~Kumar and O.~Akarsu, \emph{Bianchi type {{II}} models in the presence of perfect fluid and anisotropic dark energy}, \href{https://doi.org/10.1140/epjp/i2012-12064-4}{\emph{Eur. Phys. J. Plus} {\bfseries 127} (2012) 64} [\href{https://arxiv.org/abs/1110.2408}{{\ttfamily 1110.2408}}].

\bibitem{appleby_probing_2013}
S.A.~Appleby and E.V.~Linder, \emph{Probing {{Dark Energy Anisotropy}}}, \href{https://doi.org/10.1103/PhysRevD.87.023532}{\emph{Phys. Rev. D} {\bfseries 87} (2013) 023532} [\href{https://arxiv.org/abs/1210.8221}{{\ttfamily 1210.8221}}].

\bibitem{akarsu_scalar_2020}
O.~Akarsu, N.~Katirci, A.A.~Sen and J.A.~Vazquez, \emph{Scalar field emulator via anisotropically deformed vacuum energy: {{Application}} to dark energy}, \href{https://doi.org/10.48550/arXiv.2004.14863}{\emph{arXiv e-prints} (2020) } [\href{https://arxiv.org/abs/2004.14863}{{\ttfamily 2004.14863}}].

\bibitem{mukhanov_theory_1992}
V.~Mukhanov, \emph{Theory of cosmological perturbations}, \href{https://doi.org/10.1016/0370-1573(92)90044-Z}{\emph{Phys. Rept.} {\bfseries 215} (1992) 203}.

\bibitem{damour_non-linear_2002}
T.~Damour, I.I.~Kogan and A.~Papazoglou, \emph{Non-linear bigravity and cosmic acceleration}, \href{https://doi.org/10.1103/PhysRevD.66.104025}{\emph{Phys. Rev. D} {\bfseries 66} (2002) 104025} [\href{https://arxiv.org/abs/hep-th/0206044}{{\ttfamily hep-th/0206044}}].

\bibitem{kunz_dark_2007}
M.~Kunz and D.~Sapone, \emph{Dark {{Energy}} versus {{Modified Gravity}}}, \href{https://doi.org/10.1103/PhysRevLett.98.121301}{\emph{Phys. Rev. Lett.} {\bfseries 98} (2007) 121301} [\href{https://arxiv.org/abs/astro-ph/0612452}{{\ttfamily astro-ph/0612452}}].

\bibitem{saltas_anisotropic_2011}
I.D.~Saltas and M.~Kunz, \emph{Anisotropic stress and stability in modified gravity models}, \href{https://doi.org/10.1103/PhysRevD.83.064042}{\emph{Phys. Rev. D} {\bfseries 83} (2011) 064042} [\href{https://arxiv.org/abs/1012.3171}{{\ttfamily 1012.3171}}].

\bibitem{kolatt_constraints_2001}
T.S.~Kolatt and O.~Lahav, \emph{Constraints on {{Cosmological Anisotropy}} out to z=1 from {{Supernovae Ia}}}, \href{https://doi.org/10.1046/j.1365-8711.2001.04262.x}{\emph{Mon. Not. Roy. Astron. Soc.} {\bfseries 323} (2001) 859} [\href{https://arxiv.org/abs/astro-ph/0008041}{{\ttfamily astro-ph/0008041}}].

\bibitem{jain_search_2007}
P.~Jain, M.S.~Modgil and J.P.~Ralston, \emph{Search for {{Global Metric Anisotropy}} in {{Type Ia Supernova Data}}}, \href{https://doi.org/10.1142/S0217732307023389}{\emph{Mod. Phys. Lett. A} {\bfseries 22} (2007) 1153} [\href{https://arxiv.org/abs/astro-ph/0510803}{{\ttfamily astro-ph/0510803}}].

\bibitem{cooray_measuring_2010}
A.~Cooray, D.E.~Holz and R.~Caldwell, \emph{Measuring dark energy spatial inhomogeneity with supernova data}, \href{https://doi.org/10.1088/1475-7516/2010/11/015}{\emph{JCAP} {\bfseries 11} (2010) 015} [\href{https://arxiv.org/abs/0812.0376}{{\ttfamily 0812.0376}}].

\bibitem{brunthaler_vlbi_2010}
A.~Brunthaler, I.~{Marti-Vidal}, K.M.~Menten, M.J.~Reid, C.~Henkel, G.C.~Bower et~al., \emph{{{VLBI}} observations of {{SN}} 2008iz: {{I}}. {{Expansion}} velocity and limits on anisotropic expansion}, \href{https://doi.org/10.1051/0004-6361/201014133}{\emph{Astron. Astrophys.} {\bfseries 516} (2010) A27} [\href{https://arxiv.org/abs/1003.4665}{{\ttfamily 1003.4665}}].

\bibitem{colin_probing_2011}
J.~Colin, R.~Mohayaee, S.~Sarkar and A.~Shafieloo, \emph{Probing the anisotropic local universe and beyond with {{SNe Ia}} data}, \href{https://doi.org/10.1111/j.1365-2966.2011.18402.x}{\emph{Mon. Not. Roy. Astron. Soc.} {\bfseries 414} (2011) 264} [\href{https://arxiv.org/abs/1011.6292}{{\ttfamily 1011.6292}}].

\bibitem{cai_direction_2012}
R.~Cai and Z.~Tuo, \emph{Direction {{Dependence}} of the {{Deceleration Parameter}}}, \href{https://doi.org/10.1088/1475-7516/2012/02/004}{\emph{JCAP} {\bfseries 02} (2012) 004} [\href{https://arxiv.org/abs/1109.0941}{{\ttfamily 1109.0941}}].

\bibitem{feindt_measuring_2013}
U.~Feindt, M.~Kerschhaggl, M.~Kowalski, G.~Aldering, P.~Antilogus, C.~Aragon et~al., \emph{Measuring cosmic bulk flows with {{Type Ia Supernovae}} from the {{Nearby Supernova Factory}}}, \href{https://doi.org/10.1051/0004-6361/201321880}{\emph{Astron. Astrophys.} {\bfseries 560} (2013) A90} [\href{https://arxiv.org/abs/1310.4184}{{\ttfamily 1310.4184}}].

\bibitem{bahr-kalus_constraints_2013}
B.~{Bahr-Kalus}, D.J.~Schwarz, M.~Seikel and A.~Wiegand, \emph{Constraints on anisotropic cosmic expansion from supernovae}, \href{https://doi.org/10.1051/0004-6361/201220928}{\emph{Astron. Astrophys.} {\bfseries 553} (2013) A56} [\href{https://arxiv.org/abs/1212.3691}{{\ttfamily 1212.3691}}].

\bibitem{appleby_testing_2014}
S.~Appleby and A.~Shafieloo, \emph{Testing {{Local Anisotropy Using}} the {{Method}} of {{Smoothed Residuals I}} - {{Methodology}}}, \href{https://doi.org/10.1088/1475-7516/2014/03/007}{\emph{JCAP} {\bfseries 03} (2014) 007} [\href{https://arxiv.org/abs/1312.3415}{{\ttfamily 1312.3415}}].

\bibitem{schucker_bianchi_2014}
T.~Schucker, A.~Tilquin and G.~Valent, \emph{Bianchi {{I}} meets the {{Hubble}} diagram}, \href{https://doi.org/10.1093/mnras/stu1656}{\emph{Mon. Not. Roy. Astron. Soc.} {\bfseries 444} (2014) 2820} [\href{https://arxiv.org/abs/1405.6523}{{\ttfamily 1405.6523}}].

\bibitem{wang_probing_2014}
J.S.~Wang and F.Y.~Wang, \emph{Probing the anisotropic expansion from supernovae and {{GRBs}} in a model-independent way}, \href{https://doi.org/10.1093/mnras/stu1279}{\emph{Mon. Not. Roy. Astron. Soc.} {\bfseries 443} (2014) 1680} [\href{https://arxiv.org/abs/1406.6448}{{\ttfamily 1406.6448}}].

\bibitem{appleby_probing_2015}
S.~Appleby, A.~Shafieloo and A.~Johnson, \emph{Probing bulk flow with nearby {{SNe Ia}} data}, \href{https://doi.org/10.1088/0004-637X/801/2/76}{\emph{Astrophys. J.} {\bfseries 801} (2015) 76} [\href{https://arxiv.org/abs/1410.5562}{{\ttfamily 1410.5562}}].

\bibitem{jimenez_anisotropic_2015}
J.B.~Jimenez, V.~Salzano and R.~Lazkoz, \emph{Anisotropic expansion and {{SNIa}}: An open issue}, \href{https://doi.org/10.1016/j.physletb.2014.12.031}{\emph{Phys. Lett. B} {\bfseries 741} (2015) 168} [\href{https://arxiv.org/abs/1402.1760}{{\ttfamily 1402.1760}}].

\bibitem{javanmardi_probing_2015}
B.~Javanmardi, C.~Porciani, P.~Kroupa and J.~{Pflamm-Altenburg}, \emph{Probing the isotropy of cosmic acceleration traced by {{Type Ia}} supernovae}, \href{https://doi.org/10.1088/0004-637X/810/1/47}{\emph{Astrophys. J.} {\bfseries 810} (2015) 47} [\href{https://arxiv.org/abs/1507.07560}{{\ttfamily 1507.07560}}].

\bibitem{ghodsi_supernovae_2017}
H.~Ghodsi, S.~Baghram and F.~Habibi, \emph{Supernovae anisotropy power spectrum}, \href{https://doi.org/10.1088/1475-7516/2017/10/017}{\emph{JCAP} {\bfseries 10} (2017) 017} [\href{https://arxiv.org/abs/1609.08012}{{\ttfamily 1609.08012}}].

\bibitem{andrade_isotropy_2018}
U.~Andrade, C.A.P.~Bengaly, J.S.~Alcaniz and B.~Santos, \emph{Isotropy of low redshift type {{Ia Supernovae}}: {{A Bayesian}} analysis}, \href{https://doi.org/10.1103/PhysRevD.97.083518}{\emph{Phys. Rev. D} {\bfseries 97} (2018) 083518} [\href{https://arxiv.org/abs/1711.10536}{{\ttfamily 1711.10536}}].

\bibitem{andrade_model-independent_2018}
U.~Andrade, C.A.P.~Bengaly, B.~Santos and J.S.~Alcaniz, \emph{A {{Model-independent Test}} of {{Cosmic Isotropy}} with {{Low-z Pantheon Supernovae}}}, \href{https://doi.org/10.3847/1538-4357/aadb90}{\emph{Astrophys. J.} {\bfseries 865} (2018) 119} [\href{https://arxiv.org/abs/1806.06990}{{\ttfamily 1806.06990}}].

\bibitem{soltis_percent-level_2019}
J.~Soltis, A.~Farahi, D.~Huterer and C.M.~Liberato, \emph{Percent-{{Level Test}} of {{Isotropic Expansion Using Type Ia Supernovae}}}, \href{https://doi.org/10.1103/PhysRevLett.122.091301}{\emph{Phys. Rev. Lett.} {\bfseries 122} (2019) 091301} [\href{https://arxiv.org/abs/1902.07189}{{\ttfamily 1902.07189}}].

\bibitem{colin_evidence_2019}
J.~Colin, R.~Mohayaee, M.~Rameez and S.~Sarkar, \emph{Evidence for anisotropy of cosmic acceleration}, \href{https://doi.org/10.1051/0004-6361/201936373}{\emph{Astron. Astrophys.} {\bfseries 631} (2019) L13} [\href{https://arxiv.org/abs/1808.04597}{{\ttfamily 1808.04597}}].

\bibitem{salehi_are_2020}
A.~Salehi, H.~Farajollahi, M.~Motahari, P.~Pashamokhtari, M.~Yarahmadi and S.~Fathi, \emph{Are {{Type Ia}} supernova powerful tool to detect anisotropic expansion of the {{Universe}}?}, \href{https://doi.org/10.1140/epjc/s10052-020-8269-z}{\emph{Eur. Phys. J. C} {\bfseries 80} (2020) 753}.

\bibitem{mohayaee_supernovae_2021}
R.~Mohayaee, M.~Rameez and S.~Sarkar, \emph{Do supernovae indicate an accelerating universe?}, \href{https://doi.org/10.1140/epjs/s11734-021-00199-6}{\emph{Eur. Phys. J. ST} {\bfseries 230} (2021) 2067} [\href{https://arxiv.org/abs/2106.03119}{{\ttfamily 2106.03119}}].

\bibitem{cowell_potential_2023}
J.A.~Cowell, S.~Dhawan and H.J.~Macpherson, \emph{Potential signature of a quadrupolar hubble expansion in {{Pantheon}}+supernovae}, \href{https://doi.org/10.1093/mnras/stad2788}{\emph{Mon. Not. Roy. Astron. Soc.} {\bfseries 526} (2023) 1482} [\href{https://arxiv.org/abs/2212.13569}{{\ttfamily 2212.13569}}].

\bibitem{rahman_new_2022}
W.~Rahman, R.~Trotta, S.S.~Boruah, M.J.~Hudson and D.A.~{van Dyk}, \emph{New {{Constraints}} on {{Anisotropic Expansion}} from {{Supernovae Type Ia}}}, \href{https://doi.org/10.1093/mnras/stac1223}{\emph{Mon. Not. Roy. Astron. Soc.} {\bfseries 514} (2022) 139} [\href{https://arxiv.org/abs/2108.12497}{{\ttfamily 2108.12497}}].

\bibitem{Kalbouneh:2022tfw}
B.~Kalbouneh, C.~Marinoni and J.~Bel, \emph{{Multipole expansion of the local expansion rate}}, \href{https://doi.org/10.1103/PhysRevD.107.023507}{\emph{Phys. Rev. D} {\bfseries 107} (2023) 023507} [\href{https://arxiv.org/abs/2210.11333}{{\ttfamily 2210.11333}}].

\bibitem{Maartens:2023tib}
R.~Maartens, J.~Santiago, C.~Clarkson, B.~Kalbouneh and C.~Marinoni, \emph{{Covariant cosmography: the observer-dependence of the Hubble parameter}}, \href{https://doi.org/10.1088/1475-7516/2024/09/070}{\emph{JCAP} {\bfseries 09} (2024) 070} [\href{https://arxiv.org/abs/2312.09875}{{\ttfamily 2312.09875}}].

\bibitem{Kalbouneh:2024szq}
B.~Kalbouneh, C.~Marinoni and R.~Maartens, \emph{{Cosmography of the local Universe by multipole analysis of the expansion rate fluctuation field}}, \href{https://doi.org/10.1088/1475-7516/2024/09/069}{\emph{JCAP} {\bfseries 09} (2024) 069} [\href{https://arxiv.org/abs/2401.12291}{{\ttfamily 2401.12291}}].

\bibitem{Kalbouneh:2024yjj}
B.~Kalbouneh, J.~Santiago, C.~Marinoni, R.~Maartens, C.~Clarkson and M.~Sarma, \emph{{Expanding covariant cosmography of the local Universe: incorporating the snap and axial symmetry}}, {\emph{arXiv e-prints} (2024) } [\href{https://arxiv.org/abs/2408.04333}{{\ttfamily 2408.04333}}].

\bibitem{1984MNRAS.206..377E}
G.F.R.~{Ellis} and J.E.~{Baldwin}, \emph{{On the expected anisotropy of radio source counts}}, \href{https://doi.org/10.1093/mnras/206.2.377}{\emph{MNRAS} {\bfseries 206} (1984) 377}.

\bibitem{daSilveiraFerreira:2024ddn}
P.~da~Silveira~Ferreira and V.~Marra, \emph{{Tomographic redshift dipole: testing the cosmological principle}}, \href{https://doi.org/10.1088/1475-7516/2024/09/077}{\emph{JCAP} {\bfseries 09} (2024) 077} [\href{https://arxiv.org/abs/2403.14580}{{\ttfamily 2403.14580}}].

\bibitem{amendola_measuring_2008}
L.~Amendola, M.~Kunz and D.~Sapone, \emph{Measuring the dark side (with weak lensing)}, \href{https://doi.org/10.1088/1475-7516/2008/04/013}{\emph{JCAP} {\bfseries 04} (2008) 013} [\href{https://arxiv.org/abs/0704.2421}{{\ttfamily 0704.2421}}].

\bibitem{pitrou_weak_2013}
C.~Pitrou, J.-P.~Uzan and T.S.~Pereira, \emph{Weak lensing {{B-modes}} on all scales as a probe of local isotropy}, \href{https://doi.org/10.1103/PhysRevD.87.043003}{\emph{Phys. Rev. D} {\bfseries 87} (2013) 043003} [\href{https://arxiv.org/abs/1203.6029}{{\ttfamily 1203.6029}}].

\bibitem{schneider_weak_2006}
P.~Schneider, \emph{Weak {{Gravitational Lensing}}},  in \emph{Gravitational {{Lensing}}: {{Strong}}, {{Weak}} and {{Micro}}}, vol.~33, pp.~269--451 (2006), \href{https://doi.org/10.1007/978-3-540-30310-7_3}{DOI} [\href{https://arxiv.org/abs/astro-ph/0509252}{{\ttfamily astro-ph/0509252}}].

\bibitem{bartelmann_weak_2016}
M.~Bartelmann and M.~Maturi, \emph{Weak gravitational lensing},  Dec., 2016 [\href{https://arxiv.org/abs/1612.06535}{{\ttfamily 1612.06535}}].

\bibitem{kitching_limits_2017}
T.D.~Kitching, J.~Alsing, A.F.~Heavens, R.~Jimenez, J.D.~McEwen and L.~Verde, \emph{The {{Limits}} of {{Cosmic Shear}}}, \href{https://doi.org/10.1093/mnras/stx1039}{\emph{Mon. Not. Roy. Astron. Soc.} {\bfseries 469} (2017) 2737} [\href{https://arxiv.org/abs/1611.04954}{{\ttfamily 1611.04954}}].

\bibitem{bernardeau_full-sky_2010}
F.~Bernardeau, C.~Bonvin and F.~Vernizzi, \emph{Full-sky lensing shear at second order}, \href{https://doi.org/10.1103/PhysRevD.81.083002}{\emph{Phys. Rev. D} {\bfseries 81} (2010) 083002} [\href{https://arxiv.org/abs/0911.2244}{{\ttfamily 0911.2244}}].

\bibitem{cooray_second_2002}
A.~Cooray and W.~Hu, \emph{Second {{Order Corrections}} to {{Weak Lensing}} by {{Large-Scale Structure}}}, \href{https://doi.org/10.1086/340892}{\emph{Astrophys. J.} {\bfseries 574} (2002) 19} [\href{https://arxiv.org/abs/astro-ph/0202411}{{\ttfamily astro-ph/0202411}}].

\bibitem{hilbert_ray-tracing_2009}
S.~Hilbert, J.~Hartlap, S.D.M.~White and P.~Schneider, \emph{Ray-tracing through the {{Millennium Simulation}}: {{Born}} corrections and lens-lens coupling in cosmic shear and galaxy-galaxy lensing}, \href{https://doi.org/10.1051/0004-6361/200811054}{\emph{Astron. Astrophys.} {\bfseries 499} (2009) 31} [\href{https://arxiv.org/abs/0809.5035}{{\ttfamily 0809.5035}}].

\bibitem{schneider_b-modes_2002}
P.~Schneider, L.~{van Waerbeke} and Y.~Mellier, \emph{B-modes in cosmic shear from source redshift clustering}, \href{https://doi.org/10.1051/0004-6361:20020626}{\emph{Astron. Astrophys.} {\bfseries 389} (2002) 729} [\href{https://arxiv.org/abs/astro-ph/0112441}{{\ttfamily astro-ph/0112441}}].

\bibitem{crittenden_discriminating_2002}
R.G.~Crittenden, P.~Natarajan, U.-L.~Pen and T.~Theuns, \emph{Discriminating weak lensing from intrinsic spin correlations using the curl-gradient decomposition}, \href{https://doi.org/10.1086/338838}{\emph{Astrophys. J.} {\bfseries 568} (2002) 20} [\href{https://arxiv.org/abs/astro-ph/0012336}{{\ttfamily astro-ph/0012336}}].

\bibitem{troxel_intrinsic_2014}
M.A.~Troxel and M.~Ishak, \emph{The {{Intrinsic Alignment}} of {{Galaxies}} and its {{Impact}} on {{Weak Gravitational Lensing}} in an {{Era}} of {{Precision Cosmology}}}, \href{https://doi.org/10.1016/j.physrep.2014.11.001}{\emph{Phys. Rept.} {\bfseries 558} (2014) 1} [\href{https://arxiv.org/abs/1407.6990}{{\ttfamily 1407.6990}}].

\bibitem{chisari_intrinsic_2015}
N.E.~Chisari, S.~Codis, C.~Laigle, Y.~Dubois, C.~Pichon, J.~Devriendt et~al., \emph{Intrinsic alignments of galaxies in the {{Horizon-AGN}} cosmological hydrodynamical simulation}, \href{https://doi.org/10.1093/mnras/stv2154}{\emph{Mon. Not. Roy. Astron. Soc.} {\bfseries 454} (2015) 2736} [\href{https://arxiv.org/abs/1507.07843}{{\ttfamily 1507.07843}}].

\bibitem{codis_intrinsic_2015}
S.~Codis, R.~Gavazzi, Y.~Dubois, C.~Pichon, K.~Benabed, V.~Desjacques et~al., \emph{Intrinsic alignment of simulated galaxies in the cosmic web: Implications for weak lensing surveys}, \href{https://doi.org/10.1093/mnras/stv231}{\emph{Mon. Not. Roy. Astron. Soc.} {\bfseries 448} (2015) 3391} [\href{https://arxiv.org/abs/1406.4668}{{\ttfamily 1406.4668}}].

\bibitem{yamauchi_weak_2012}
D.~Yamauchi, T.~Namikawa and A.~Taruya, \emph{Weak lensing generated by vector perturbations and detectability of cosmic strings}, \href{https://doi.org/10.1088/1475-7516/2012/10/030}{\emph{JCAP} {\bfseries 10} (2012) 030} [\href{https://arxiv.org/abs/1205.2139}{{\ttfamily 1205.2139}}].

\bibitem{yamauchi_full-sky_2013}
D.~Yamauchi, T.~Namikawa and A.~Taruya, \emph{Full-sky formulae for weak lensing power spectra from total angular momentum method}, \href{https://doi.org/10.1088/1475-7516/2013/08/051}{\emph{JCAP} {\bfseries 08} (2013) 051} [\href{https://arxiv.org/abs/1305.3348}{{\ttfamily 1305.3348}}].

\bibitem{pitrou_weak-lensing_2015}
C.~Pitrou, T.S.~Pereira and J.-P.~Uzan, \emph{Weak-lensing by the large scale structure in a spatially anisotropic universe: Theory and predictions}, \href{https://doi.org/10.1103/PhysRevD.92.023501}{\emph{Phys. Rev. D} {\bfseries 92} (2015) 023501} [\href{https://arxiv.org/abs/1503.01125}{{\ttfamily 1503.01125}}].

\bibitem{pereira_weak-lensing_2016}
T.S.~Pereira, C.~Pitrou and J.-P.~Uzan, \emph{Weak-lensing {$B$}-modes as a probe of the isotropy of the universe}, \href{https://doi.org/10.1051/0004-6361/201527258}{\emph{Astron. Astrophys.} {\bfseries 585} (2016) L3} [\href{https://arxiv.org/abs/1503.01127}{{\ttfamily 1503.01127}}].

\bibitem{kitching_3d_2014}
{\scshape CFHTLenS} collaboration, \emph{{{3D Cosmic Shear}}: {{Cosmology}} from {{CFHTLenS}}}, \href{https://doi.org/10.1093/mnras/stu934}{\emph{Mon. Not. Roy. Astron. Soc.} {\bfseries 442} (2014) 1326} [\href{https://arxiv.org/abs/1401.6842}{{\ttfamily 1401.6842}}].

\bibitem{blanchard_euclid_2020}
{\scshape Euclid} collaboration, \emph{Euclid preparation: {{VII}}. {{Forecast}} validation for {{Euclid}} cosmological probes}, \href{https://doi.org/10.1051/0004-6361/202038071}{\emph{Astron. Astrophys.} {\bfseries 642} (2020) A191} [\href{https://arxiv.org/abs/1910.09273}{{\ttfamily 1910.09273}}].

\bibitem{deshpande_euclid_2024}
{\scshape Euclid} collaboration, \emph{Euclid preparation: {{XXVIII}}. {{Modelling}} of the weak lensing angular power spectrum}, \href{https://doi.org/10.1051/0004-6361/202346110}{\emph{Astron. Astrophys.} {\bfseries 684} (2024) A138} [\href{https://arxiv.org/abs/2302.04507}{{\ttfamily 2302.04507}}].

\bibitem{smith_stable_2003}
{\scshape VIRGO Consortium} collaboration, \emph{Stable clustering, the halo model and nonlinear cosmological power spectra}, \href{https://doi.org/10.1046/j.1365-8711.2003.06503.x}{\emph{Mon. Not. Roy. Astron. Soc.} {\bfseries 341} (2003) 1311} [\href{https://arxiv.org/abs/astro-ph/0207664}{{\ttfamily astro-ph/0207664}}].

\bibitem{takahashi_revising_2012}
R.~Takahashi, M.~Sato, T.~Nishimichi, A.~Taruya and M.~Oguri, \emph{Revising the {{Halofit Model}} for the {{Nonlinear Matter Power Spectrum}}}, \href{https://doi.org/10.1088/0004-637X/761/2/152}{\emph{Astrophys. J.} {\bfseries 761} (2012) 152} [\href{https://arxiv.org/abs/1208.2701}{{\ttfamily 1208.2701}}].

\bibitem{bird_massive_2012}
S.~Bird, M.~Viel and M.G.~Haehnelt, \emph{Massive {{Neutrinos}} and the {{Non-linear Matter Power Spectrum}}}, \href{https://doi.org/10.1111/j.1365-2966.2011.20222.x}{\emph{Mon. Not. Roy. Astron. Soc.} {\bfseries 420} (2012) 2551} [\href{https://arxiv.org/abs/1109.4416}{{\ttfamily 1109.4416}}].

\bibitem{pitrou_theory_2007}
C.~Pitrou, T.S.~Pereira and J.-P.~Uzan, \emph{Theory of cosmological perturbations in an anisotropic universe}, \href{https://doi.org/10.1088/1475-7516/2007/09/006}{\emph{JCAP} {\bfseries 09} (2007) 006} [\href{https://arxiv.org/abs/0707.0736}{{\ttfamily 0707.0736}}].

\bibitem{aghanim_planck_2020}
{\scshape Planck} collaboration, \emph{Planck 2018 results. {{VI}}. {{Cosmological}} parameters}, \href{https://doi.org/10.1051/0004-6361/201833910}{\emph{Astron. Astrophys.} {\bfseries 641} (2020) A6} [\href{https://arxiv.org/abs/1807.06209}{{\ttfamily 1807.06209}}].

\bibitem{workman_review_2022}
{\scshape Particle Data Group} collaboration, \emph{Review of {{Particle Physics}}}, \href{https://doi.org/10.1093/ptep/ptac097}{\emph{PTEP} {\bfseries 2022} (2022) 083C01}.

\bibitem{almeida_structure_2022}
J.P.B.~Almeida, J.~{Motoa-Manzano}, J.~Nore{\~n}a, T.S.~Pereira and C.A.~{Valenzuela-Toledo}, \emph{Structure formation in an anisotropic universe: {{Eulerian}} perturbation theory}, \href{https://doi.org/10.1088/1475-7516/2022/02/018}{\emph{JCAP} {\bfseries 02} (2022) 018} [\href{https://arxiv.org/abs/2111.06756}{{\ttfamily 2111.06756}}].

\bibitem{fleury_light_2015}
P.~Fleury, \emph{Light Propagation in Inhomogeneous and Anisotropic Cosmologies}, Ph.D. thesis, Universit{\'e} Pierre-et-Marie-Curie, Nov., 2015.
\newblock \href{https://arxiv.org/abs/1511.03702}{{\ttfamily 1511.03702}}.

\bibitem{pontzen_linearization_2011}
A.~Pontzen and A.~Challinor, \emph{Linearization of homogeneous, nearly-isotropic cosmological models}, \href{https://doi.org/10.1088/0264-9381/28/18/185007}{\emph{Class. Quant. Grav.} {\bfseries 28} (2011) 185007} [\href{https://arxiv.org/abs/1009.3935}{{\ttfamily 1009.3935}}].

\bibitem{pitrou_bianchi_2019}
C.~Pitrou and T.S.~Pereira, \emph{Bianchi spacetimes as super-curvature modes around isotropic cosmologies}, \href{https://doi.org/10.1103/PhysRevD.100.123534}{\emph{Phys. Rev. D} {\bfseries 100} (2019) 123534} [\href{https://arxiv.org/abs/1909.13688}{{\ttfamily 1909.13688}}].

\bibitem{abramo_testing_2010}
L.R.~Abramo and T.S.~Pereira, \emph{Testing gaussianity, homogeneity and isotropy with the cosmic microwave background}, \href{https://doi.org/10.1155/2010/378203}{\emph{Adv. Astron.} {\bfseries 2010} (2010) 378203} [\href{https://arxiv.org/abs/1002.3173}{{\ttfamily 1002.3173}}].

\bibitem{samandar_cosmic_2024}
{\scshape COMPACT} collaboration, \emph{Cosmic topology. {{Part IIIa}}. {{Microwave}} background parity violation without parity-violating microphysics}, \href{https://doi.org/10.48550/arXiv.2407.09400}{\emph{arXiv e-prints} (2024) } [\href{https://arxiv.org/abs/2407.09400}{{\ttfamily 2407.09400}}].

\bibitem{hajian_measuring_2003}
A.~Hajian and T.~Souradeep, \emph{Measuring statistical isotropy of the {{CMB}} anisotropy}, \href{https://doi.org/10.1086/379757}{\emph{Astrophys. J. Lett.} {\bfseries 597} (2003) L5} [\href{https://arxiv.org/abs/astro-ph/0308001}{{\ttfamily astro-ph/0308001}}].

\bibitem{hajian_statistical_2003}
A.~Hajian and T.~Souradeep, \emph{Statistical isotropy of {{CMB}} and cosmic topology}, {\emph{arXiv e-prints} (2003) } [\href{https://arxiv.org/abs/astro-ph/0301590}{{\ttfamily astro-ph/0301590}}].

\bibitem{hajian_cosmic_2005}
A.~Hajian and T.~Souradeep, \emph{The {{Cosmic}} microwave background bipolar power spectrum: {{Basic}} formalism and applications}, {\emph{arXiv e-prints} (2005) } [\href{https://arxiv.org/abs/astro-ph/0501001}{{\ttfamily astro-ph/0501001}}].

\bibitem{ghosh_unveiling_2007}
T.~Ghosh, A.~Hajian and T.~Souradeep, \emph{Unveiling {{Hidden Patterns}} in {{CMB Anisotropy Maps}}}, \href{https://doi.org/10.1103/PhysRevD.75.083007}{\emph{Phys. Rev. D} {\bfseries 75} (2007) 083007} [\href{https://arxiv.org/abs/astro-ph/0604279}{{\ttfamily astro-ph/0604279}}].

\bibitem{joshi_bipolar_2010}
N.~Joshi, S.~Jhingan, T.~Souradeep and A.~Hajian, \emph{Bipolar {{Harmonic}} encoding of {{CMB}} correlation patterns}, \href{https://doi.org/10.1103/PhysRevD.81.083012}{\emph{Phys. Rev. D} {\bfseries 81} (2010) 083012} [\href{https://arxiv.org/abs/0912.3217}{{\ttfamily 0912.3217}}].

\bibitem{book_odd-parity_2012}
L.G.~Book, M.~Kamionkowski and T.~Souradeep, \emph{Odd-{{Parity Bipolar Spherical Harmonics}}}, \href{https://doi.org/10.1103/PhysRevD.85.023010}{\emph{Phys. Rev. D} {\bfseries 85} (2012) 023010} [\href{https://arxiv.org/abs/1109.2910}{{\ttfamily 1109.2910}}].

\bibitem{aluri_novel_2015}
P.K.~Aluri, N.~Pant, A.~Rotti and T.~Souradeep, \emph{A novel approach to reconstructing signals of isotropy violation from a masked {{CMB}} sky}, \href{https://doi.org/10.1103/PhysRevD.92.083015}{\emph{Phys. Rev. D} {\bfseries 92} (2015) 083015} [\href{https://arxiv.org/abs/1506.00550}{{\ttfamily 1506.00550}}].

\bibitem{kilbinger_precision_2017}
M.~Kilbinger, C.~Heymans, M.~Asgari, S.~Joudaki, P.~Schneider, P.~Simon et~al., \emph{Precision calculations of the cosmic shear power spectrum projection}, \href{https://doi.org/10.1093/mnras/stx2082}{\emph{Mon. Not. Roy. Astron. Soc.} {\bfseries 472} (2017) 2126} [\href{https://arxiv.org/abs/1702.05301}{{\ttfamily 1702.05301}}].

\bibitem{challinor_weak_2006}
A.~Challinor and A.~Lewis, \emph{Weak {{Gravitational Lensing}} of the {{CMB}}}, \href{https://doi.org/10.1016/j.physrep.2006.03.002}{\emph{Phys. Rept.} {\bfseries 429} (2006) 1} [\href{https://arxiv.org/abs/astro-ph/0601594}{{\ttfamily astro-ph/0601594}}].

\bibitem{challinor_lensed_2005}
A.~Challinor and A.~Lewis, \emph{Lensed {{CMB}} power spectra from all-sky correlation functions}, \href{https://doi.org/10.1103/PhysRevD.71.103010}{\emph{Phys. Rev. D} {\bfseries 71} (2005) 103010} [\href{https://arxiv.org/abs/astro-ph/0502425}{{\ttfamily astro-ph/0502425}}].

\bibitem{lesgourgues_cosmic_2011-3}
J.~Lesgourgues, T.~Tram and D.~Blas, \emph{The {{Cosmic Linear Anisotropy Solving System}} ({{CLASS}}) {{II}}: {{Approximation}} schemes}, \href{https://doi.org/10.1088/1475-7516/2011/07/034}{\emph{JCAP} {\bfseries 07} (2011) 034} [\href{https://arxiv.org/abs/1104.2933}{{\ttfamily 1104.2933}}].

\bibitem{hu_power_1999}
W.~Hu, \emph{Power {{Spectrum Tomography}} with {{Weak Lensing}}}, \href{https://doi.org/10.1086/312210}{\emph{Astrophys. J. Lett.} {\bfseries 522} (1999) L21} [\href{https://arxiv.org/abs/astro-ph/9904153}{{\ttfamily astro-ph/9904153}}].

\bibitem{harris_array_2020}
C.R.~Harris, K.J.~Millman, S.J.~van~der Walt, R.~Gommers, P.~Virtanen, D.~Cournapeau et~al., \emph{Array {{Programming}} with {{NumPy}}}, \href{https://doi.org/10.1038/s41586-020-2649-2}{\emph{Nature} {\bfseries 585} (2020) 357} [\href{https://arxiv.org/abs/2006.10256}{{\ttfamily 2006.10256}}].

\bibitem{virtanen_scipy_2020}
P.~Virtanen, R.~Gommers, T.E.~Oliphant, M.~Haberland, T.~Reddy, D.~Cournapeau et~al., \emph{{{SciPy}} 1.0--{{Fundamental Algorithms}} for {{Scientific Computing}} in {{Python}}}, \href{https://doi.org/10.1038/s41592-019-0686-2}{\emph{Nature Meth.} {\bfseries 17} (2020) 261} [\href{https://arxiv.org/abs/1907.10121}{{\ttfamily 1907.10121}}].

\bibitem{hunter_matplotlib_2007}
J.D.~Hunter, \emph{Matplotlib: {{A 2D Graphics Environment}}}, \href{https://doi.org/10.1109/MCSE.2007.55}{\emph{Comput. Sci. Eng.} {\bfseries 9} (2007) 90}.

\bibitem{arfken_mathematical_2005}
G.B.~Arfken and H.-J.~Weber, \emph{Mathematical Methods for Physicists}, Elsevier, Boston, 6th ed~ed. (2005).

\bibitem{edmonds_angular_1996}
A.R.~Edmonds, \emph{Angular Momentum in Quantum Mechanics}, Princeton Landmarks in Physics, Princeton University Press, Princeton (1996).

\bibitem{durrer_cosmic_2020}
R.~Durrer, \emph{The Cosmic Microwave Background}, Cambridge University Press, New York, second edition~ed. (2020).

\bibitem{newman_note_1966}
E.T.~Newman and R.~Penrose, \emph{Note on the {{Bondi-Metzner-Sachs Group}}}, \href{https://doi.org/10.1063/1.1931221}{\emph{J. Math. Phys.} {\bfseries 7} (1966) 863}.

\bibitem{goldberg_spin_1967}
J.N.~Goldberg, A.J.~Macfarlane, E.T.~Newman, F.~Rohrlich and E.C.G.~Sudarshan, \emph{Spin- {\emph{s}} {{Spherical Harmonics}} and {\dh}}, \href{https://doi.org/10.1063/1.1705135}{\emph{J. Math. Phys.} {\bfseries 8} (1967) 2155}.

\bibitem{boyle_how_2016}
M.~Boyle, \emph{How should spin-weighted spherical functions be defined?}, \href{https://doi.org/10.1063/1.4962723}{\emph{J. Math. Phys.} {\bfseries 57} (2016) 092504} [\href{https://arxiv.org/abs/1604.08140}{{\ttfamily 1604.08140}}].

\bibitem{hu_weak_2000}
W.~Hu, \emph{Weak {{Lensing}} of the {{CMB}}: {{A Harmonic Approach}}}, \href{https://doi.org/10.1103/PhysRevD.62.043007}{\emph{Phys. Rev. D} {\bfseries 62} (2000) 043007} [\href{https://arxiv.org/abs/astro-ph/0001303}{{\ttfamily astro-ph/0001303}}].

\end{thebibliography}\endgroup
